\documentclass[12pt]{article}

\newenvironment{wileykeywords}{\textsf{Keywords:}\hspace{\stretch{1}}}{\hspace{\stretch{1}}\rule{1ex}{1ex}}

\usepackage{amsmath,amssymb}
\usepackage{graphicx}
\usepackage{setspace}
\usepackage[normal]{caption}
\usepackage{color}
\usepackage{dcolumn}
\usepackage{bm}
\usepackage[numbers,super,comma,sort&compress]{natbib}

\usepackage{enumerate}
\usepackage{multirow}
\usepackage{threeparttable}
\usepackage{longtable}

\definecolor{background-color}{gray}{0.98}

\newcommand{\abs}[1]{\left| #1 \right|} 
\let\baraccent=\= 
\renewcommand{\=}[1]{\stackrel{#1}{=}} 

\newcommand{\E}[1]{\left< #1 \right>} 

\usepackage[colorlinks=true,citecolor=blue,linkcolor=blue]{hyperref}

\title{Alchemical Grid Dock (AlGDock): Binding Free Energy Calculations between Flexible Ligands and Rigid Receptors}
\author{David D. L. Minh \thanks{Department of Chemistry, Illinois Institute of Technology, Chicago, IL 60616, USA}}

\begin{document}

\maketitle

\begin{abstract}
Alchemical Grid Dock (AlGDock) is open-source software designed to compute the binding potential of mean force (BPMF) - the binding free energy between a flexible ligand and a rigid receptor - for a small organic ligand and a biological macromolecule. Multiple BPMFs can be used to rigorously compute binding affinities between flexible partners. AlGDock uses replica exchange between thermodynamic states at different temperatures and receptor-ligand interaction strengths. Receptor-ligand interaction energies are represented by interpolating precomputed grids. Thermodynamic states are adaptively initialized and adjusted on-the-fly to maintain replica exchange rates. In demonstrative calculations, when the bound ligand is treated as fully solvated, AlGDock estimates BPMFs with a precision within 4 kT in 65\% and within 8 kT for 91\% of systems. It correctly identifies the native binding pose in 83\% of simulations. Performance is sometimes limited by subtle differences in the important configuration space of sampled and targeted thermodynamic states.
\end{abstract}

\begin{wileykeywords}
Protein-Ligand, Noncovalent Binding Free Energy, 
Implicit Ligand Theory, Thermodynamic Length, Replica Exchange
\end{wileykeywords}

\clearpage




\begin{figure}[p]
\centering
\colorbox{background-color}{
\fbox{
\begin{minipage}{1.0\textwidth}
\includegraphics[width=2in]{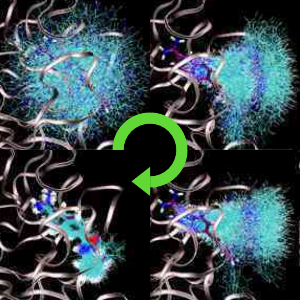} \\
Configurations sampled from representative thermodynamic states as the temperature is reduced and receptor-ligand interactions are scaled in. These samples are used to estimate binding free energies between flexible small molecules and rigid receptors. These estimates can be used for estimating binding free energies between flexible binding partners.
\end{minipage}
}}
\end{figure}

  \makeatletter
  \renewcommand\@biblabel[1]{#1.}
  \makeatother

\bibliographystyle{apsrev}

\renewcommand{\baselinestretch}{1.5}
\normalsize

\clearpage

\section*{\sffamily \Large INTRODUCTION} 


Alchemical Grid Dock (AlGDock) is an open-source computer program designed to compute 
the binding potential of mean force (BPMF) -
the binding free energy between a flexible ligand and a rigid receptor -
between a small organic ligand and a biological macromolecule.

The BPMF is defined as a ratio of configurational integrals \cite{Minh2012},
\begin{eqnarray}
B(r_R) &=& -\beta^{-1} \ln \left( 
\frac{ \int  I(\xi) J(\xi) e^{-\beta U(r_{RL})} ~ dr_L d\xi}
       { \int  I(\xi) J(\xi) e^{-\beta [U(r_{L}) + U(r_R)] } ~ dr_L d\xi } \right).
\label{eq:BPMF}
\end{eqnarray}
In this paper, the 
internal coordinates (excluding translation and rotation) 
of a receptor-ligand complex, $r_{RL}$, are partitioned 
into the receptor, $r_R$,
the ligand, $r_L$,
and the relative translation and rotation of the species, $\xi$.
$\beta = (k_B T)^{-1}$ is the inverse of Boltzmann's constant times the temperature.
$I(\xi)$ is an indicator function that specifies whether the receptor and ligand are bound (1) or not (0).
$J(\xi)$ is the Jacobian for transforming Cartesian coordinates into the coordinate system used for $r_L$ and $\xi$.
$U(\cdot)$ is the potential energy of a species in solvent.

BPMFs are useful for characterizing noncovalent association processes.
According to implicit ligand theory (ILT) \cite{Minh2012,Nguyen2018}, 
the standard binding free energy can be computed 
from BPMFs between a ligand and multiple receptor conformations.
Moreover, ILT explains that BPMFs can reweight receptor conformations 
from the apo (ligand-free) to the holo (ligand-bound) ensemble for the ligand of interest.
This approach has the greatest potential benefit when computing binding free energies 
or averages over respective holo ensembles for many ligands to a single receptor;
after performing receptor sampling once, the same snapshots may be used for many ligands.
Another potential use for BPMFs is as a secondary scoring function for molecular docking.

A number of BPMF calculations have been published in the scientific literature.
In the first paper on ILT, BPMFs were estimated for simple host-guest systems \cite{Minh2012}.
Unlike the calculations herein, the calculations in this first paper 
did not employ computational shortcuts that exploit the rigidity of the receptor.
Without necessarily referring to their calculations as BPMFs, 
several groups have computed binding free energies between simple ligands 
and rigid conformations of the protein T4 lysozyme:
\citet{Mobley2007} deployed rigorous alchemical binding free energy calculations;
\citet{Ucisik2014} developed a fast and approximate method;
and my research group published a study 
where BPMFs were computed using AlGDock \cite{Xie2017} 
and another where they were computed based on 
a fast Fourier Transform \cite{Nguyen2018a}.
In my group's former publications involving AlGDock, 
we cited an unpublished earlier version of this article \cite{Minh2015}.
In \citet{Xie2017}, we computed standard binding free energies for 141 ligands using multiple BPMFs
and showed that our results accurately 
reproduce values from flexible-receptor simulations for 25 ligands.
The same BPMFs were used to demonstrate 
our new formalism for estimating relative, opposed to absolute, binding free energies \cite{Nguyen2018}.
In the Drug Design Data Resource Grand Challenge 3, 
a blinded challenge for binding affinity and pose prediction, 
my research group submitted entries based on BPMFs \cite{Xie2019}.
For one system, vascular endothelial growth factor receptor 2, 
our submissions were among the most highly correlated with experiment.
Results from our purely physics-based approach were competitive 
with methods using knowledge-based potentials, which were the best performers in the challenge.
As discussed in \citet{Xie2019}, several issues led to weaker performance in other subchallenges,
including the neglect of DMSO and SO$_4$ in the binding site of Cathepsin S and 
poor selection of receptor snapshots for free energy calculations with Janus Kinase 2 and Mitogen-activated protein kinase 14.
Based on its uneven performance in the challenge, it appears that BPMF calculations show potential 
but further development is necessary before broader use.
In this paper, I describe BPMF calculations 
for a variety of protein receptors:
the Astex diverse set \cite{Hartshorn2007}, 
a curated database of 85 high-quality crystallographic structures 
of protein-ligand complexes with pharmaceutical or agrochemical interest.

AlGDock uses methods that are similar to those used in recent alchemical binding free energy calculations with a flexible receptor \cite{Gallicchio2012, Wang2013b, Wang2015} and also implements algorithms that make BPMF calculations faster and more robust.
As in other work, AlGDock performs Boltzmann sampling for a series of thermodynamic states
with different degrees of coupling between the receptor and ligand 
and periodically attempts Monte Carlo moves to exchange configurations between the different replicas.
The main methodological distinctions are (1) the use of precomputed nonbonded interaction grids for receptor-ligand interactions \cite{Pattabiraman1985, Meng1992, Minh2018} and (2) the adaptive initialization and on-the-fly adjustment of thermodynamic states.
The former accelerates BPMFs compared to flexible-receptor binding free energy calculations because evaluating nonbonded terms no longer scales as $O(N^2)$ (neglecting cutoffs) with the number of receptor atoms $N$.
Rather, once the grid is computed, calculation time does not depend on $N$.
The latter improves the precision of free energy estimates by ensuring sufficient configuration space overlap between adjacent thermodynamic states along the alchemical protocol.

AlGDock is a python module 
based on the Molecular Modeling Toolkit (MMTK) 2.7.8 \cite{Hinsen2000}. 
It is available under the open-source MIT license at \url{https://github.com/ccbatiit/algdock/}.

\section*{\sffamily \Large METHODOLOGY}




This section details the algorithms in AlGDock and describes demonstrative BPMF calculations for the Astex diverse set \cite{Hartshorn2007}. Parameter values specified below, e.g. the binding site radius, were used in the demonstrative calculations, but most are adjustable arguments to the program.

\section*{\sffamily \Large Thermodynamic Cycle}

\begin{figure}[p]
\begin{center}
\includegraphics[width=3.18in]{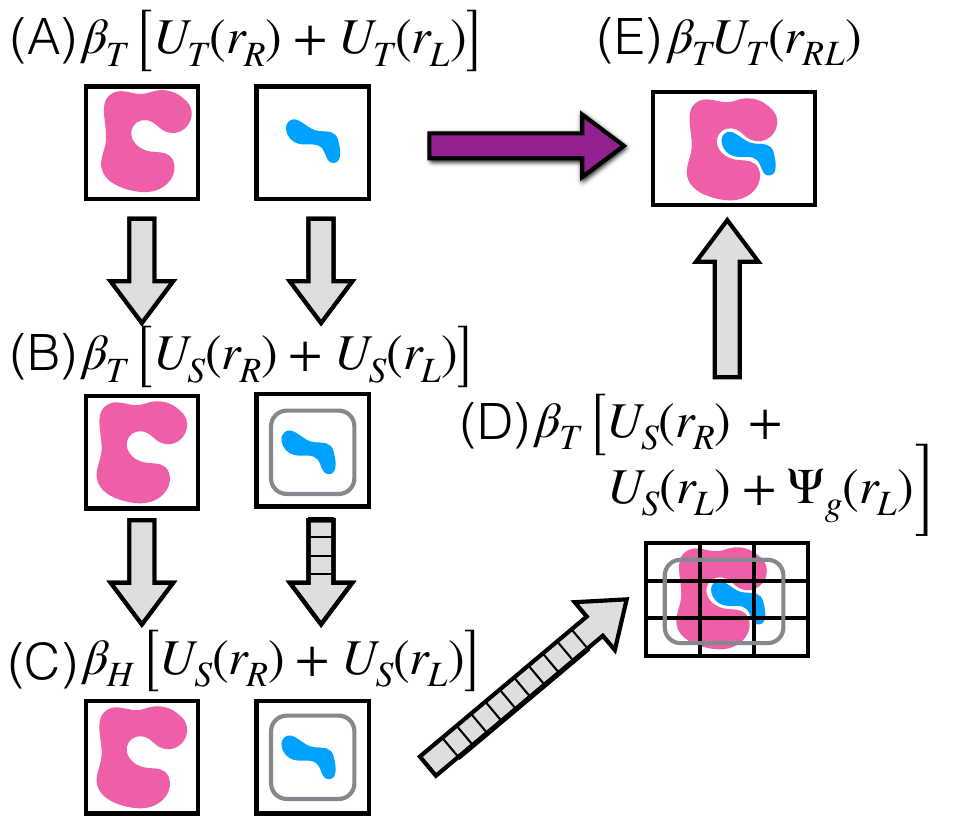}
\end{center}
\caption{
{\bf Thermodynamic cycle for BPMFs.} 
Milestone thermodynamic states are labeled with letters in parentheses and expressions for the reduced potential energy.
$\beta_T^{-1} = k_B (300 \textrm{~K})$ and $\beta_H^{-1} = k_B (600 \textrm{~K})$ are inverse temperature factors for the target and high temperatures, respectively.
$U_T(\cdot)$ and $U_S(\cdot)$ denote potential energies for the target and sampling force fields, respectively.
These potential energies include 
molecular mechanics terms and the implicit solvent model.
$\Psi_g(\cdot)$ is the potential energy due to receptor-ligand interaction grids.
Arrows with orthogonal lines indicate multiple intermediate thermodynamic states. For BPMF calculations, configurations are sampled from thermodynamic states with the rounded boxes and from their intermediates.
\label{fig:thermo_cycle}}
\end{figure}

In AlGDock, BPMFs are calculated based on the thermodynamic cycle shown in Figure \ref{fig:thermo_cycle}.
Figure \ref{fig:thermo_cycle} shows milestone thermodynamic states, which are referred to with the letters A to E.
The states between and including milestones X and Y will be referred to as states XY.

Over the course of this cycle, the receptor-ligand interaction strength is scaled and the temperature is varied.
The temperature is varied because high-temperature states enhance transitions between local energetic minima.
As this paper will deal with thermodynamic states at different temperatures,
I will frequently refer to the \emph{reduced potential energy} \cite{Shirts2008}, 
a log probability density that incorporates the inverse temperature factor $\beta = (k_BT)^{-1}$.
Reduced potential energies for key milestones in the thermodynamic cycle are shown in Figure \ref{fig:thermo_cycle}.
Subsequently, 
the reduced potential energy will be denoted with a lowercase $u$.
Furthermore, the reduced free energy difference between two milestones $X$ and $Y$ will be denoted as $f_{XY}$.
Because converting from reduced to standard potential energies and free energies involves dividing by $\beta$, 
the units of these reduced quantities are $k_B T$.

In all of the simulated thermodynamic states, the ligand is confined to the binding site using a flat-bottom harmonic potential \cite{Gallicchio2012,Wang2013b,Minh2012},
\begin{eqnarray}
	u_I(d) = \left\{
		\begin{array}{lcc}
			0 & \textrm{if} & d \leq d_0 \\
			\frac{1}{2}\beta k(d-d_o)^2 & \textrm{if} & d > d_0 \\
		\end{array}
	\right.,
\label{eq:binding_site}
\end{eqnarray}
where $k = 10000$ kJ/(mol nm$^2$) is the spring constant, $d$ is the distance between the ligand center of mass and the center of the binding site, and $d_0 = 6.0$ \AA~is the radius of the binding site. There is no restriction on ligand rotation.

The thermodynamic cycle involves sampling and target force fields, which may be distinct from each other.
In the demonstrative calculations, the sampling and target force fields had much in common, but some important distinctions.
They both used the AMBER ff14SB force field for proteins and ions and Generalized Amber Force Field 2 \cite{Wang2004a} with AM1BCC charges \cite{Jakalian1999,Jakalian2002} for other molecules.
The key differences between the target and sampling force fields were the solvation model and whether receptor-ligand interactions were evaluated directly or by grid interpolation.
They were also implemented via different python modules.
In the sampling force field, AlGDock uses MMTK (which was extended to include implicit solvent and grid interpolation terms) to calculate energies and forces.
In contrast, the target force field was evaluated with OpenMM \cite{Eastman2010}.
The sampling force field uses the generalized Born/surface area model II from \citet{Onufriev2004} (OBC), adapted from OpenMM \cite{Eastman2010}, as an implicit solvent model.
It differs from the target force field because only the ligand, opposed to the entire complex, is assumed to be solvated, and because it is sometimes scaled down, as described in the next paragraph.
Finally, the sampling force field models receptor-ligand interactions with grid interpolation, as described further below.
In the target force field, these interaction energies are directly computed.

To elaborate on the implicit solvent model, the sampling force field in AlGDock can employ two solvation pathways: Desolvated and Full. 
In the Desolvated pathway, an implicit solvent model for the ligand is present in milestone B but its strength is linearly scaled down and is zero at milestone C.
Implicit solvent is not used for states CD, 
saving computer time.
This pathway makes the most sense 
if the bound ligand is nearly completely desolvated.
In the Full solvation pathway, the implicit solvent is at full strength for states BD.
This pathway makes the most sense if the bound ligand 
is nearly fully solvated.
In either case, an implicit solvent model for the complex (opposed to just the ligand) is used in milestones A and E.
Therefore, the final results for either solvation pathway should be equivalent 
in the limit of asymptotic sampling,
but may be distinct for incompletely converged calculations.

The grid interaction energy,
\begin{eqnarray}
\Psi_g(r_{RL}) = \Psi_{PBSA}(r_{RL}) + \Psi_{vdW}(r_{RL}),
\end{eqnarray}
is based on one electrostatic and two van der Waals grids.

The electrostatic interaction energy $\Psi_{PBSA}(r_{RL})$ is evaluated by multiplying atomic partial charges with the electrostatic potential.
The electrostatic potential for each ligand configuration is obtained by trilinear interpolation of a precomputed grid.
The grid is produced by solving the linear Poisson-Boltzmann equation around the minimized receptor molecule using APBS 1.4 \cite{Baker2001} with sequential focusing.
Coarse grids are at least 1.5 times larger than the range of the receptor molecule in each dimension.  Fine grids have the same size as the van der Waals grids, and a spacing of 0.5 \AA.
Coarse grids use multiple Debye-Huckel boundary conditions, and fine grids use coarse-grid solutions as boundary conditions.
Both grids are solved with the following options: 
a quintic B-spline charge discretization, spline window width of 0.3, protein dielectric of 2.0, solvent dielectric of 80.0, solvent density of 10.0, solvent radius of 1.4 \AA, smoothed dielectric and ion-accessibility coefficients, and temperature of 300.0 K. 

The van der Waals interaction energy $\Psi_{vdW}(r_{RL})$ is evaluated by an analogous grid-based procedure \cite{Minh2018}.
This procedure is built on the ideas of \citet{Pattabiraman1985} and \citet{Meng1992}, who precomputed van der Waals energies at positions along a grid.
To account for the highly nonlinear nature of van der Waals potentials \cite{Oberlin1998},
energies are calculated using a transformation, trilinear interpolation, and inverse transformation \cite{Venkatachalam2003}.
Based on my previous recommendation \cite{Minh2018}, an inverse transformation power of 4 is used for the repulsive potential and no transformation for the attractive potential.
(A reasonable alternative approach could be the logarithmic interpolation proposed by \citet{Diller1999}.)

\begin{figure}[p]
\begin{center}
\includegraphics[width=3.18in]{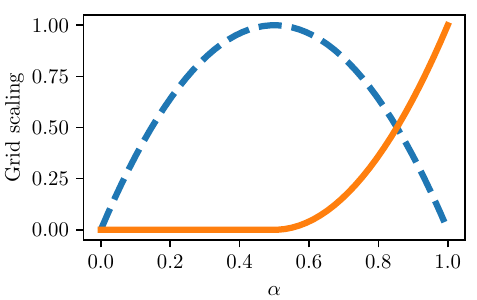}
\end{center}
\caption{
{\bf Grid scaling for states CD.} $\alpha_{sg}$ (dashed line) and $\alpha_g$ (solid line) as a function of the progress variable $\alpha$.
\label{fig:grid_scaling}}
\end{figure}

Alchemical transformations that modulate the strength of interactions between atoms often face an ``end-point catastrophe'' in which free energy changes are numerically unstable \cite{Michel2010}.
The end-point catastrophe occurs because steric overlaps that do not lead to high energies when molecules are decoupled can do so when coupling is added.
To circumvent this issue, a set of soft Lennard-Jones repulsive and electrostatic grids is introduced between milestones C and D. In these soft grids, the original grid value, $v_{o}$, is replaced with $v_{max} \tanh \left( v_o/v_{max} \right)$. (\citet{Gallicchio2012} also used a hyperbolic tangent energy cap.) For the soft Lennard-Jones repulsive grid, $v_{max} = 10.0$ kJ mol$^{-1/2}$. 
A potential issue with soft Lennard-Jones is that they can be overwhelmed by electrostatic contributions.
To circumvent this issue of electrostatic pinning, the soft electrostatic grid uses as maximum value such that the electrostatic energy is less than or equal to the soft Lennard-Jones repulsive energy for every heavy atom at every grid point. This is established by setting $v_{max}$ for the electrostatic grid to 10 times the minimum ratio of Lennard-Jones and electrostatic scaling factors. The reduced potential energy is switched according to the protocol,
\begin{eqnarray}
u_\alpha(r_{RL}) &=& \frac{1}{k_B T(\alpha)} \left[U_s(r_L) + U_s(r_R) + \alpha_{sg}(\alpha) \Psi_{sg}(r_{RL}) + \alpha_{g}(\alpha) \Psi_g(r_{RL}) \right] \\
\alpha_{sg}(\alpha) &=& -\left(2 \alpha - 1\right)^2 + 1 \nonumber \\
\alpha_{g}(\alpha) &=& \frac{\left(2 \alpha - 1\right)^2}{1 + \exp \left[-1000 (\alpha - \frac{1}{2}) \right]} \nonumber \\
T(\alpha) &=& (T_T - T_H) \alpha + T_H \nonumber 
\end{eqnarray}
This protocol turns on the soft grids first, and then the unperturbed grids (Figure \ref{fig:grid_scaling}).
The potential is consistent with milestone C at $\alpha=0$ and milestone D at $\alpha=1$.

\section*{\sffamily \Large Sampling}

For states BD, ligand conformational sampling is performed with a combination of Hamiltonian Monte Carlo (HMC) \cite{Duane1987}, external coordinate Markov chain Monte Carlo moves, and Hamiltonian replica exchange. 

HMC trial moves are based on 50 steps of velocity Verlet molecular dynamics with an adaptive time step that ranges between 0.1 and 5.0 fs. Time step adaptation occurs during the initialization of thermodynamic states, as described in the next section.

To accelerate transitions between binding poses, external coordinate Markov chain Monte Carlo moves are attempted for states CD when $\alpha < 0.01$. The external coordinate move consists of:
\begin{enumerate}
\item Random rotation. A random quaternion is converted into a matrix that is used to rotate the molecule about its center of mass. 
\item Random translation. The magnitude of translation in each dimension is drawn from a Gaussian distribution with a standard deviation of 0.6 \AA.
\end{enumerate}
The move is accepted or rejected according to the Metropolis criterion. Moves are not attempted for $\alpha > 0.01$ due to low acceptance probability.

During production, Hamiltonian replica exchange \cite{Jiang2010,Gallicchio2012,Wang2013b,Minh2012} 
is attempted for states BC and states CD.
Replica exchange is a Markov chain Monte Carlo move that swaps the configurations of a pair of simulations at different thermodynamic states. (Equivalently, it may be regarded as swapping the states.) 
Consider the thermodynamic states $a$ and $b$ with reduced energies $u_a$ and $u_b$, respectively. 
If $x$ is the original configuration in state $a$ and $y$ the original configuration in state $b$, then the acceptance probability,
\begin{eqnarray}
p_{acc} &=& \min \left[ 1, e^{-u_a(y)-u_b(x)+u_a(x)+u_b(y)} \right],
\label{eq:repXacc}
\end{eqnarray}
preserves the Boltzmann distribution in both states.

Typical replica exchange protocols attempt exchanges between pairs of neighboring thermodynamic states, but this restriction is unnecessary. 
As replica exchange is a type of Gibbs sampling \cite{Chodera2011a}, an arbitrary number of attempts can be made between arbitrary pairs of states.
In AlGDock, each sweep of replica exchange includes attempts to swap configurations between pairs of states that are $1, 2, ..., \min(5, K)$ states apart, where $K$ is the total number of thermodynamic states in the direction. 

\section*{\sffamily \Large Stages}

BPMF calculations with AlGDock are broken down into the following stages:
\begin{enumerate}
\item \textbf{Ligand preparation:} the ligand is minimized with 5000 steepest descent steps. The temperature is ramped from 20 K to 300 K over 30 geometrically spaced simulations of 2500 steps each.
\item \textbf{Initialization:} starting from 50 seed configurations, simulations of 2000 steps are used to initialize each thermodynamic state for states BD.
\item \textbf{Equilibration and production:} simulations are conducted for states BC and subsequently for states CD.
\item \textbf{Postprocessing:} samples from milestones B and D are postprocessed using the target force field.
\item \textbf{Estimation:} Free energy differences that sum up to the BPMF are estimated.
\end{enumerate}

\subsection*{\sffamily Initialization}

The purpose of initialization is to establish a protocol with reasonable time steps and mean replica exchange rate $\E{p_{acc}}$ between all neighboring states.
The key benefit of replica exchange is to spread sampled configurations across a range of different thermodynamic states.
A bottleneck in the exchange of configurations across the pair of states can eliminate this benefit of replica exchange; 
groups of thermodynamic states separated by the bottleneck effectively become independent.
Low exchange probabilities are also indicative of poor configuration space overlap,
which can limit the convergence of free energy estimates \citep{Lu2001}.

For states BC, 
the first thermodynamic state ($k=0$) is at 300 K and the ligand is steadily warmed to 600 K.
The first thermodynamic state initialized for states CD depends on whether 
there is a fully-bound pose available in the binding site.
If a pose is available, then the first state is fully bound at 300 K.
Otherwise, the first state is fully unbound at 600 K.
In the latter situation, 50 randomly selected configurations from milestone C are placed in the binding site at 453 random center-of-mass positions (0.5 positions per \AA$^3$) and rotated with 100 different random orientations.

After state $k$ is initialized, state $k+1$ is initialized as follows:
\begin{enumerate}
\item \textbf{Parameter selection:} Parameters for state $k+1$ are selected using a new algorithm designed to separate states at approximately even intervals in thermodynamic length.

Thermodynamic length is a metric of the distance on the manifold of thermodynamic states \cite{Weinhold1975}. For a sequence of states, the statistical error in free energy calculations is minimized and the replica exchange frequency is nearly maximized when intermediate states are equidistant in thermodynamic length \cite{Shenfeld2009}.
Suppose that thermodynamic parameters (e.g. temperature, pressure, grid scaling strengths) are specified by a vector $\mathbf{\lambda}$ with components $\lambda^i$.
Let $\gamma \equiv \gamma(\alpha)$ describe the dependence of $\mathbf{\lambda}$ on the variable $\alpha$, such that $\gamma(0)$ is the initial and $\gamma(1)$ is the final thermodynamic state.
For microscopic systems, the thermodynamic length is defined by the path integral \cite{Crooks2007, Shenfeld2009},
\begin{eqnarray}
\mathcal L & \equiv & \int_0^1 \sqrt{\sum_{i,j} \frac{\partial \gamma^i}{\partial \alpha} g(\gamma)_{ij} \frac{\partial \gamma^j}{\partial \alpha} } d\alpha.
\end{eqnarray}
Given a parameter vector $\lambda$, the reduced potential energy is $u_\lambda(x) = U_\lambda(x)/(k_B T_\lambda)$, where $U_\lambda(x)$ is the effective potential energy and $T_\lambda$ is the temperature. The normalized log probability of observing a configuration $x$ is $l_\lambda(x) = -u_\lambda(x) - \ln Z_\lambda$, where $Z_\lambda = \int e^{-u_\lambda(x)} dx$ is the partition function. These quantities are used to define elements of the Fisher information matrix,
\begin{eqnarray}
g(\gamma)_{ij} \equiv \sigma^2_\lambda \left[ \partial^i l_\lambda, \partial^j l_\lambda \right],
\end{eqnarray}
where $\sigma^2_\lambda$ is the covariance in state $\lambda$ and $\partial^i$ denotes a partial derivative with respect to $\lambda^i$. 
For a protocol in which only one parameter $\lambda^i$ varies with $\alpha$, the length is,
$\mathcal L = \int_0^1 \frac{\partial \gamma^i}{\partial \alpha} \sigma_\lambda \left[ \partial^i l_\lambda \right] dt$.

Numerical estimates of $\mathcal L$ are most accurate when samples are drawn from many states between $0<\alpha<1$ \cite{Crooks2007}. Such exhaustive sampling, however, is unavailable during initialization. A simple approximation for the thermodynamic length when one parameter changes is,
$\mathcal L = \Delta \lambda^i ~ \sigma_0 \left[ \partial^i l_\lambda \right]$,
where $\Delta \lambda^i$ is the total change in the value of the parameter $\lambda^i$ and $\sigma_0 \left[ \partial^i l_\lambda \right]$ is a standard deviation in the initial state. Thus, if one desires $\mathcal L$ to be approximately constant between different intermediate stages in a protocol, then the change in parameter should be inversely proportional to $\sigma_0 \left[ \partial^i l_\lambda \right]$,
\begin{eqnarray}
\Delta \lambda^i = \frac{s}{\sigma_0 \left[ \partial^i l_\lambda \right]},
\end{eqnarray}
where $s$ is an adjustable parameter, the \emph{thermodynamic speed}. 

For example, 
with the Full solvation pathway, 
the parameter that varies between milestones B and C is the temperature, T. 
As the log probability of a ligand configuration is
$l_\lambda = -\frac{U_S(r_L)}{k_B T} - \ln Z_\lambda$,
T is incremented by,
\begin{eqnarray}
\Delta \lambda^i &=& -\frac{s_{bc} k T^2}{\sigma_\lambda[U_S]},
\end{eqnarray}
where $s_{bc} = 20.0$.
For states CD,
the log probability of $r_{RL}$ is
$l_\lambda = -u_\alpha(r_{RL}) - \ln Z_\lambda$.
$\alpha$ is incremented by,
\begin{eqnarray}
\Delta \lambda^i = s_{cd} \left[ 
\abs{\frac{d \alpha_{sg}}{d \alpha}} \frac{\sigma_\lambda[\Psi_{sg}]}{k_B T(\alpha)} 
+ \abs{\frac{d \alpha_{g}}{d \alpha}} \frac{\sigma_\lambda[\Psi_{g}]}{k_B T(\alpha)}
+ \abs{T_T - T_H} \frac{\sigma_\lambda[u_\alpha(r_{RL})]}{T(\alpha)} \right]^{-1},
\end{eqnarray}
with $s_{cd} = 0.2$.
If the targeted value of the parameter is exceeded (e.g. temperature increased above 600 K), then the targeted value is used.

\item \textbf{Seed selection:} 50 configurations from state $k$ are resampled as starting seeds for simulations in state $k+1$.

Configurations are drawn from state $k$ with weights proportional to $\exp[u_k(x_i) - u_{k+1}(x_i)]$, where $u_k(x) = U_k(x)/(k_B T_k)$ is the reduced potential energy in state $k$. In the limit of infinite sampling of state $k$, resampled configurations would be Boltzmann-distributed in state $k+1$. With imperfect sampling of state $k$, resampled configurations approximate the Boltzmann distribution in state $k+1$. The seed selection process is an example of what is known in the statistics literature as sampling importance resampling.

\item \textbf{Sampling and adaptation:} Simulations of 2000 steps are run from each seed and the sampling protocol is adapted to obtain a reasonable acceptance rate.

If the Monte Carlo acceptance rate is greater than 0.8, the time step is increased by 0.125 fs. If it is less than 0.4, then the time step is reduced by 0.25 fs. If it is less than 0.1, then the time step is reduced by 0.5 fs. Initialization is repeated at the new time step until the acceptance rate is between 0.4 and 0.8.

\item \textbf{Verification:} The mean replica exchange probability, $\E{p_{acc}}$, is used to verify that thermodynamic states are not too distinct nor similar. 

It is estimated by taking the sample mean of $p_{acc}$ (Eq. \ref{eq:repXacc}) for every pair of initial samples (at the same time index) from states $k$ and $k+1$. If $\E{p_{acc}}$ is estimated to be too low \cite{Nguyen2016} (below 0.4), then parameters for state $k+1$ are reselected with a smaller increment (thermodynamic speed is adjusted by a factor of 4/5) and simulations are repeated. If it is too high (above 0.99), then state $k$ is removed.
\end{enumerate}

\subsection*{\sffamily Equilibration and production}

Equilibration and production calculations are broken down into cycles.
Each cycle consists of 1000 iterations of the following: 
an HMC move and 20 external coordinate MCMC moves (if $\alpha < 0.01$) for each thermodynamic state 
and then 25 sweeps of replica exchange.
50 snapshots are saved per replica exchange cycle.
The demonstrative calculations are based on 8 cycles between for states BC and 15 cycles for states CD.

Between each cycle, simulation data are saved and thermodynamic states are inserted as necessary. The purpose of inserting thermodynamic states is to ensure adequate replica exchange acceptance rates. As discussed above, AlGDock includes an estimate of $\E{p_{acc}}$ to verify new thermodynamic states during initialization. However, the configuration space explored during replica exchange can be distinct from that explored during initialization, leading to a substantial change in the observed $p_{acc}$. Thus, if at the end of a cycle the average replica exchange acceptance rate between any pair of neighboring thermodynamic states falls to less than 0.4 \cite{Nguyen2016}, then another thermodynamic state is inserted between the pair. Sampling importance resampling is used to populate configurations in this new state. That is, samples for the new state are drawn from samples for other states with probability proportional to the density in the new state.

Separation of equilibration and production is based on a method inspired by \citet{Chodera2016}. The integrated autocorrelation time and statistical inefficiency is estimated based on the mean potential energy of configurations from the last $c \in \{1, 2, ..., C\}$ cycles of equilibration and production, where $C$ is the total number of cycles. The number of statistically independent samples is determined by dividing the number of snapshots in $c$ cycles by the estimated statistical inefficiency. The simulation is considered equilibrated based on the value of $c$ that provides the largest number of statistically independent samples.

\subsection*{\sffamily Estimation}

BPMFs were estimated according to 
$\beta_T B(r_R) = f_{AB} + f_{BC,L} + f_{CD}' + f_{DE}$.
$f_{BC,L}$ is the free energy of warming the ligand from $T_T = 300$ K to $T_H = 600$ K, and,
\begin{eqnarray}
f_{CD}' &=& -\ln \frac{\int I(\xi) J(\xi) e^{-\beta_T [U(r_L) + \Psi_g(r_{RL})]} dr_L d\xi}{\int I(\xi) J(\xi) e^{-\beta_H U(r_L)} dr_L d\xi}.
\end{eqnarray}
$f_{BC,L} + f_{CD}'$ is used instead of $f_{BC} + f_{CD}$ because the former calculation can be performed without determining the receptor internal energy $U(r_R)$.
The receptor desolvation free energy is estimated by the difference, $f_{AB,R} = \beta_T (U(r_R) - U(r_R))$.
Other free energy differences are estimated based on equilibrated samples from replica exchange 
for states BC or states CD.
$f_{AB,R}$ and $f_{DE}$ are estimated by free energy perturbation \cite{Zwanzig1954} using configurations drawn from milestones A and E, respectively.
$f_{BC,L}$ and $f_{CD}'$ are estimated by the multistate Bennett acceptance ratio \cite{Shirts2008}, which uses potential energies from every replica.

\subsection*{\sffamily Pose prediction}

Binding poses were predicted based on ligand conformations sampled after equilibration. 
Configurations sampled from milestone D were 
clustered using hierarchical clustering with complete linkage, 
performed using scipy.cluster.hierarchy.linkage in scipy v0.14.0 \cite{Walt2011numpy}.
Distances between snapshots were based on the 
Hungarian symmetry-corrected heavy-atom root mean square deviation (RMSD),
which is also implemented in UCSF DOCK 6 \cite{Lang2009}.
Clusters were separated based on a threshold of 1.0 \AA. 
The probability of each cluster was obtained by reweighing configurations via the factor,
\begin{equation}
w_c = \exp \left[ -\beta_T \left( U_T(r_{RL}) - U_S(r_L) - \Psi_g(r_L) \right)\right].
\label{eq:reweighing_all}
\end{equation}
or by assuming that interaction energies are the only terms that change between milestones D and E,
\begin{equation}
w_c = \exp \left[ -\beta_T \left( U_T(r_{RL}) - U_T(r_{L}) - \Psi_g(r_L) \right)\right].
\label{eq:reweighing_interact}
\end{equation}
Pose predictions were based on the lowest-energy configuration, 
according to the force field in milestone E, from each cluster.
Reduced free energies of each pose $p$ were based on the cumulative weight of configurations in the cluster,
\begin{equation}
f_{EE_p} = -\ln \frac{\sum_c w_c}{\sum_p \sum_c w_c},
\label{eq:pose_free_energy}
\end{equation}
where $w_c$ is the weight of the configuration, $\sum_c$ is over configurations in the cluster, 
and $\sum_p$ is a sum over poses.
If multiple independent simulations were run, the pose was predicted based on the pose 
with the lowest interaction energy, lowest total energy, or by the lowest pose-specific BPMF,
\begin{eqnarray}
f_{AE,p} = f_{AE} + f_{EE_p}
\end{eqnarray}

\subsection*{\sffamily Astex diverse set BPMF calculations}

For each system in the Astex diverse set \cite{Hartshorn2007}, 11 independent simulations were performed using the Desolvated and Full solvation pathways. 
Input files in AMBER format (based on the ff14SB force field for proteins and ions and Generalized Amber Force Field 2 \cite{Wang2004a} with AM1BCC charges \cite{Jakalian1999,Jakalian2002} for other molecules) were reused from a previous study \cite{Minh2018}.
Simulations were started from the crystallographic pose and poses obtained from molecular docking.
Docking with UCSF DOCK 6 \cite{Lang2009} was repeated using a similar procedure as in the previous study \cite{Minh2018}, but with a minimum anchor size of 5 instead of 40, which increases binding pose sampling.

After starting poses were minimized for 1000 conjugate gradient steps in the appropriate force field for milestone D, the lowest-energy pose was used to initialize simulations in the milestone. After the thermodynamic states were initialized, docked poses within the binding site were also used as starting points for each state in replica exchange. The lowest-energy pose was used for replica exchange with milestone D. Higher-energy poses were used in intermediate replicas to fill all available thermodynamic states. If there were more states than docked poses, the lowest-energy pose was duplicated. 

Calculations were performed the Open Science Grid \cite{Pordes2007}, supercomputing resources managed by the National Science Foundation eXtreme Science and Engineering Discovery Environment (XSEDE) \citep{xsede}, and on the Minh group computing cluster at IIT. Benchmark calculations were run using single-core jobs with standard compute nodes on XSEDE Comet of the San Diego Supercomputer Center. These nodes have Intel Xeon E5-2680v3 processors, 128 GB DDR4 DRAM (64 GB per socket), and 320 GB of SSD local scratch memory.

\section*{\sffamily \Large RESULTS}

\section*{\sffamily \Large Thermodynamic state initialization and adaptation is system-specific and robust}

\begin{figure}[p]
\begin{center}
\includegraphics[width=3.18in]{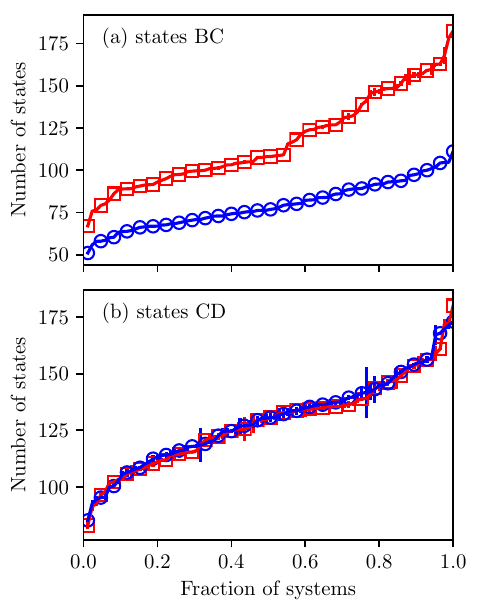}
\end{center}
\caption{
{\bf Number of thermodynamic states} (a) 
for states BC, and (b) for states CD.
The marker indicates the mean value and error bars the standard deviation of 11 independent simulations based on the Desolvated (red squares) and Full (blue circles) solvation options.
They are ordered by the mean number of states.
\label{fig:n_states}}
\end{figure}

The described method for thermodynamic state initialization and adaptation yields protocols that are tailored for specific systems.
The large range in the number of thermodynamic states, $N_{states}$, provides evidence that protocols are system-specific (Figure \ref{fig:n_states}).
For states BC, there is generally larger number of states with the Desolvated pathway (between 67 and 182) than for the Full pathway (between 51 and 111).
The larger number of states is likely because of a more significant difference between the end states with the Desolvated pathway due to the removal of implicit solvent at milestone C.
In contrast, the number of states for the Desolvated and Full pathways for states CD is comparable.

In addition to being system-specific, the protocols appear to be robust.
In all the systems, the standard deviation of the number of states, $\sigma[N_{states}]$ is small relative to the average number of states, $\bar{N}_{states}$.
For states BC, $\sigma[N_{states}]$ is less than 2 for all systems.
For states CD, the protocols are more variable.
Variability may be larger because interaction grids 
introduce more possibilities for trapping in local minima during initialization. 
For 1l7f, the protocols for state CD
appear to fall into two distinct classes (Figure S1 in the Supplementary Material).
It is worth noting, however, that higher variation in protocols does not necessarily lead to inaccurate or imprecise results.

Similarly, the initialization of each state appears to be adaptive and robust.
In the vast majority of initialization processes, the time step converged to between 2.75 and 3.75 fs (Figure S2 in the Supplementary Material). For states CD, some protocols have shorter time steps. The shortest time steps are from simulations with 1lrh. The relationship between the progress variable $\alpha$ and time step is fairly consistent across the independent protocols, demonstrating that the time step adaptation procedure is robust.

\section*{\sffamily \Large Replica exchange acceptance probabilities are reasonable}

\begin{figure}[p]
\begin{center}
\includegraphics[width=3.18in]{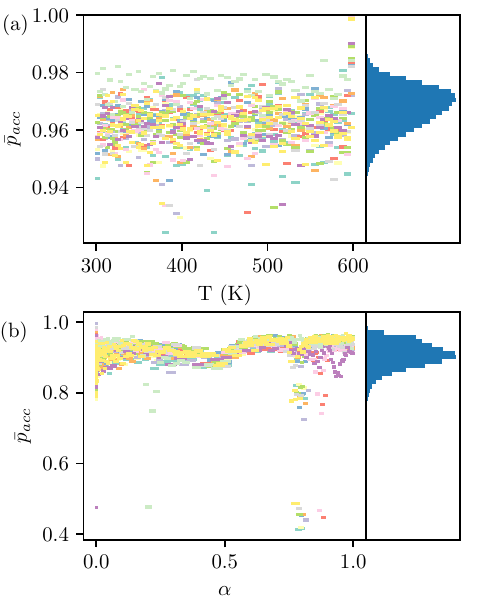}
\end{center}
\caption{
{\bf Mean acceptance probability statistics.} 
$\bar{p}_{acc}$ for fifteen protocols with the lowest observed $\bar{p}_{acc}$ 
(a) for states BC and 
(b) and states CD
are shown with the a line connecting neighboring states.
Histograms of $\bar{p}_{acc}$ from all simulations are shown on the right panel.
The largest bin count is 
14492 for states BC and 
31068 for states CD.
For states CD, the simulations are from 1opk (4), 1r1h (4), 1t40 (3), 1v48 (2),  1oq5 (1), 1jje (1), 
where the number of simulations is in the parentheses.
\label{fig:mean_acc}}
\end{figure}

Estimates of $\E{p_{acc}}$ indicate that protocols do not have any replica exchange bottlenecks (Figure \ref{fig:mean_acc}).
The notation $\bar{p}_{acc}$ refers to a statistical estimator for $\E{p_{acc}}$.
For states BC, $\bar{p}_{acc}$ estimated during replica exchange are high and have low variance. This outcome is consistent with achieving the goal of nearly equal thermodynamic length between adjacent thermodynamic states. It also implies that the configuration spaces explored during initialization and replica exchange are largely the same. For states CD, the replica exchange rates are also high but there is a larger variance. While most $\bar{p}_{acc}$ are between 0.7 and 1.0, there are a few simulations where $\bar{p}_{acc}$ is much lower. In these simulations, the acceptance probability dips around $\alpha=0.2$ or $\alpha=0.8$, but are high for most other values of $\alpha$. These drops indicate that different conformations are explored in replica exchange compared to during thermodynamic state initialization. However, even these low $\bar{p}_{acc}$ are not low enough to be considered a bottleneck; they are expected to allow configurations to pass through the thermodynamic states several times per cycle. 

The observed $\bar{p}_{acc}$ also underscore the point that large variation in protocols is not necessarily problematic. While some systems have a relatively large variation in $N_{states}$ for states BC, all $\bar{p}_{acc}$ estimated from replica exchange are 0.92 or greater.
Longer protocols simply have a higher $\bar{p}_{acc}$.

\section*{\sffamily \Large Simulation time is dependent on the solvation time and ligand size}

\begin{figure}[p]
\begin{center}
\includegraphics[width=3.18in]{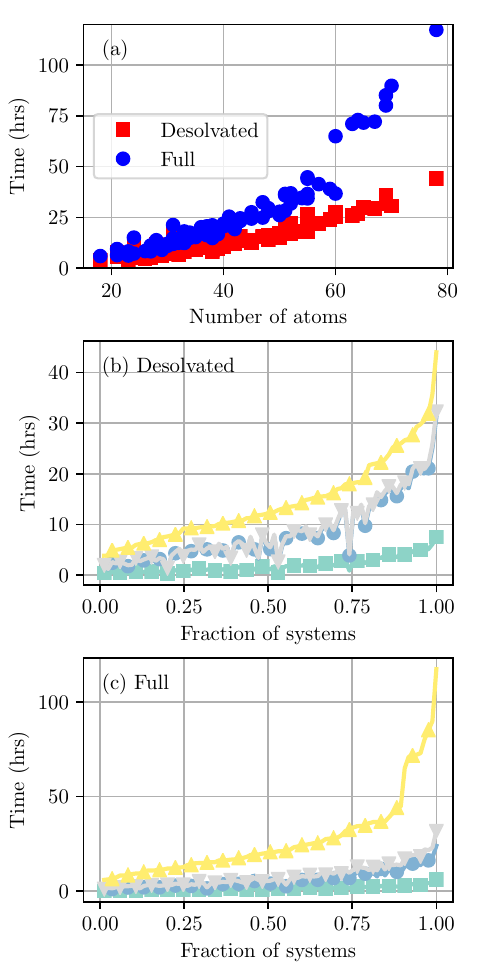}
\end{center}
\caption{
{\bf Benchmark simulation times.}
(a) Scatter plot of the total time for Desolvated (red squares) and Full (blue circles) solvation options 
as a function of the number of atoms in the system.
Breakdown of times for (b) Desolvated and (c) Full solvation options.
Benchmark simulations include initialization, equilibration and production, postprocessing, and free energy estimation.
From bottom to top, lines depict the cumulative time 
through initialization (cyan squares) and estimation (violet circles) for states AC 
and then initialization (pink downward triangles) and estimation (yellow upward triangles) for states CE.
Systems are ordered along the x axis by the total simulation time.
\label{fig:benchmark_timings}}
\end{figure}

The total CPU time of benchmark simulations spanned a large range (Figure \ref{fig:benchmark_timings}).
For simulations with the Desolvated option, they ranged from 4 to 44 hours.
For simulations with the Full solvation option, they ranged from 6 to 117 hours.

Benchmark simulation times are dependent on the solvation option and are roughly exponentially dependent on ligand size (Figure \ref{fig:benchmark_timings}a). 
Simulations with the the Full solvation option are substantially slower than those with the Desolvated option, particularly for larger ligands.
For both solvation options, the dependence appears roughly linear up to 40 atoms.
Above this threshold, the difference in computation time between solvation options becomes more pronounced.

In all simulations, the majority of time was spent in equilibration and production
(Figure \ref{fig:benchmark_timings}b and \ref{fig:benchmark_timings}c).
The largest fraction of time was states CD, and the second largest fraction for states BC. Initialization and estimation for states CD consumes a small but noticeable fraction of simulation time, whereas the analogous steps for states BC consume a nearly negligible fraction of simulation time.

\section*{\sffamily \Large Simulations sample a variety of poses}

\begin{figure}[p]
\begin{center}
\includegraphics[width=6.64in]{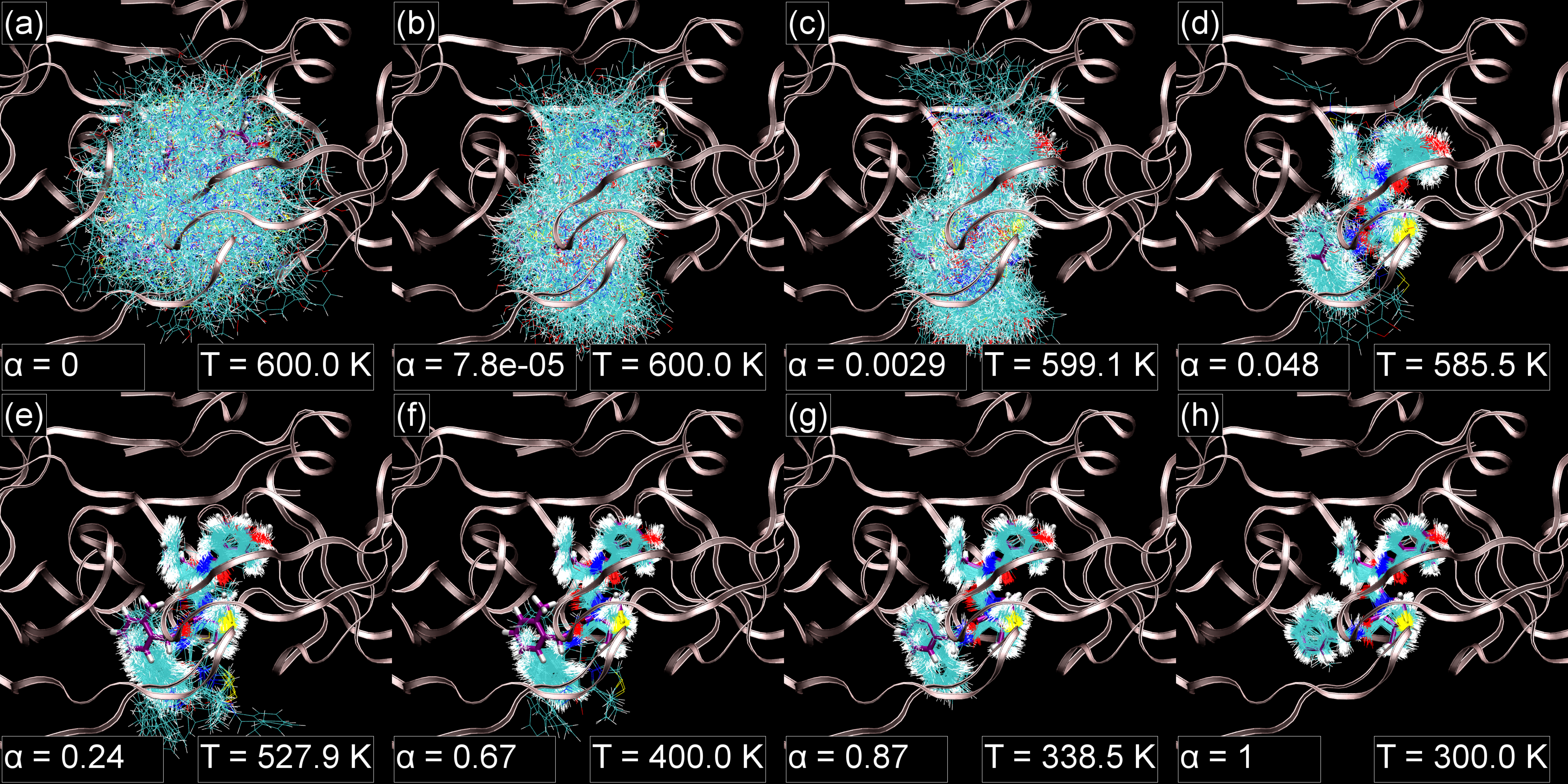}
\end{center}
\caption{
{\bf Samples from evenly spaced thermodynamic states for states CD},
taken from a representative simulation of 1kzk with the Desolvated solvation option.
The protein structure structure is shown with ribbons and 
the crystallographic ligand pose is shown with a thick licorice representation 
and purple carbon atoms. 
The same illustration scheme is used in Figures \ref{fig:intermediates_1ia1_full} and \ref{fig:dock_last}.
These figures were generated with VMD \cite{Humphrey1996}.
\label{fig:intermediates_1kzk_desolvated}}
\end{figure}

At milestone C, sampled configurations are uniformly distributed in a sphere. As $\alpha$ increases from 0 to 1 for states CD, the configuration space of the ligand is gradually restricted. First, the soft grids prevent ligand atoms from overlapping with receptor atoms. At intermediate values of $\alpha$, the ligand may assume several poses (e.g. Figure \ref{fig:intermediates_1kzk_desolvated}). Finally, at milestone D, the ligand usually but does not always sample from a single minimum.

\begin{figure}[p]
\begin{center}
\includegraphics[width=6.64in]{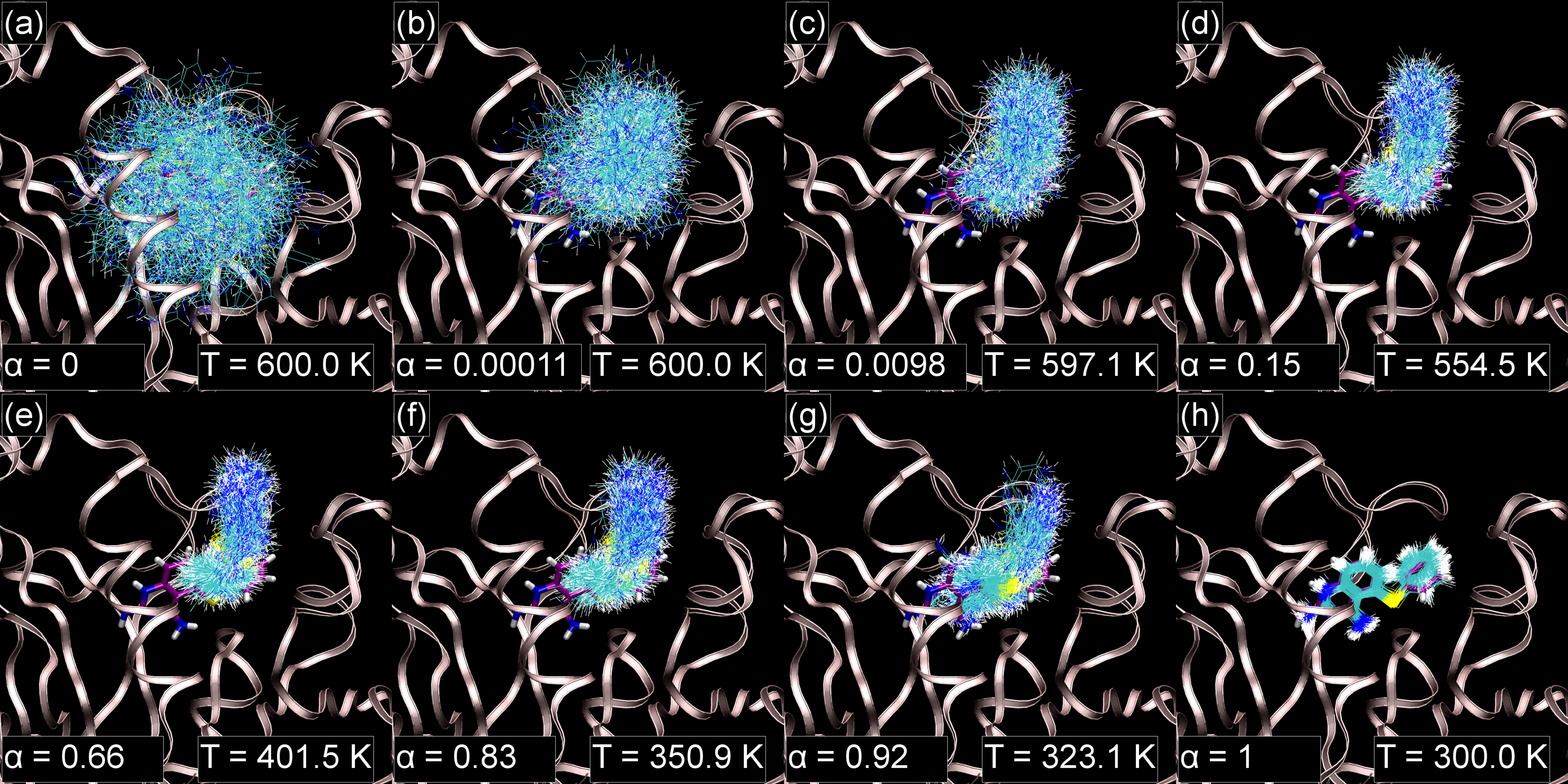}
\end{center}
\caption{
{\bf Samples from evenly spaced thermodynamic states for states CD},
taken from a representative simulation of 1l7f with the Full solvation option.
\label{fig:intermediates_1ia1_full}}
\end{figure}

In many simulations, the configuration space sampled at higher $\alpha$ is a subset of that sampled at lower $\alpha$ (e.g. Figure \ref{fig:intermediates_1kzk_desolvated}). In others, however, the conformational minima for $\alpha \approx 0.5$, where the soft grids are at full strength, are entirely distinct from the minima for $\alpha \approx 1$ (e.g. Figure \ref{fig:intermediates_1ia1_full}). When there are shifts in important configuration space, the thermodynamic states at $\alpha \approx 0.5$ are less beneficial to sampling from milestone D. Nonetheless, configuration space overlap between adjacent states is a sufficient condition for precise free energy estimates.

\begin{figure}[p]
\begin{center}
\includegraphics[width=6.64in]{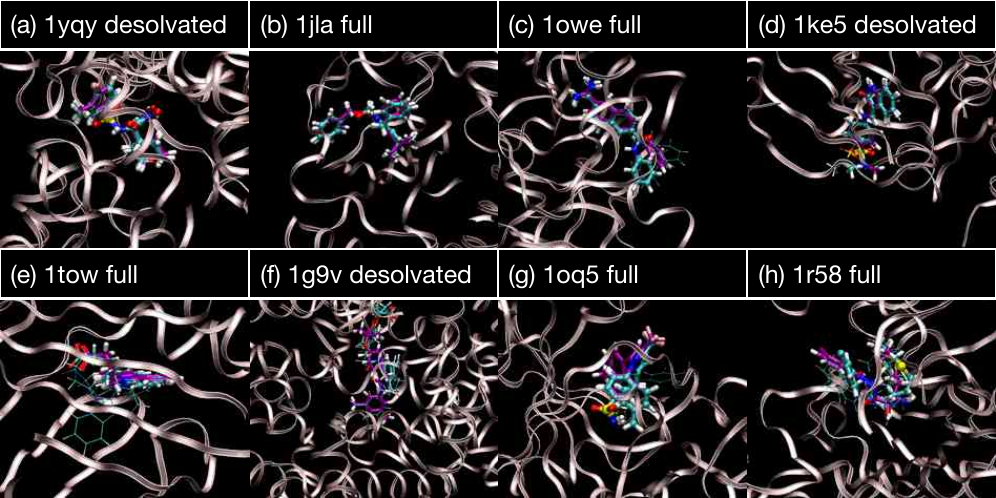}
\end{center}
\caption{
{\bf Predicted poses} taken from representative simulations. 
A licorice representation is used for both the crystallographic pose (purple carbon atoms) and predicted poses (cyan).
For the pose predictions, the thickness of the representation is proportional to its Boltzmann weight (using energies from the OBC model).
Poses are shown for which weights are greater than 0.001.
\label{fig:dock_last}}
\end{figure}

At milestone D, most simulations sample from a single minimum (Figure \ref{fig:dock_last}ab), 
but there are several other situations. These include sampling alternate poses 
that share a common warhead position, but have a floppier tail (Figure \ref{fig:dock_last}cd),
or that are clearly distinct (Figure \ref{fig:dock_last}efgh).
When there are distinct poses, 
the native pose is often correctly identified as being most populated (Figure \ref{fig:dock_last}e), 
but sometimes other poses are calculated to have greater weight (Figure \ref{fig:dock_last}fgh).
Indeed, the native pose may not even been among the predicted poses (Figure \ref{fig:dock_last}gh).

\section*{\sffamily \Large Complex force field transfer limits BPMF precision}

The mean and standard deviation of free energy differences between all adjacent pairs of milestones in the thermodynamic cycle (Figure \ref{fig:thermo_cycle}) and for the total BPMF are reported in Table S1 of the supplementary material. While BPMFs are estimated within chemical precision of 1 kcal/mol = 1.68 RT for only 28.2\% of systems, the majority of BPMFs are estimated within 4 $k_B T$ (75.3\% with Desolvated and 74.1\% with Full options) and nearly all within 8 $k_B T$ (87.1\% with Desolvated and 94.1\% with Full). The largest source of imprecision is estimation of $f_{DE}$, which is associated with transferring the complex between sampling and targeted force fields.

\begin{figure}[p]
\begin{center}
\includegraphics[width=6.64in]{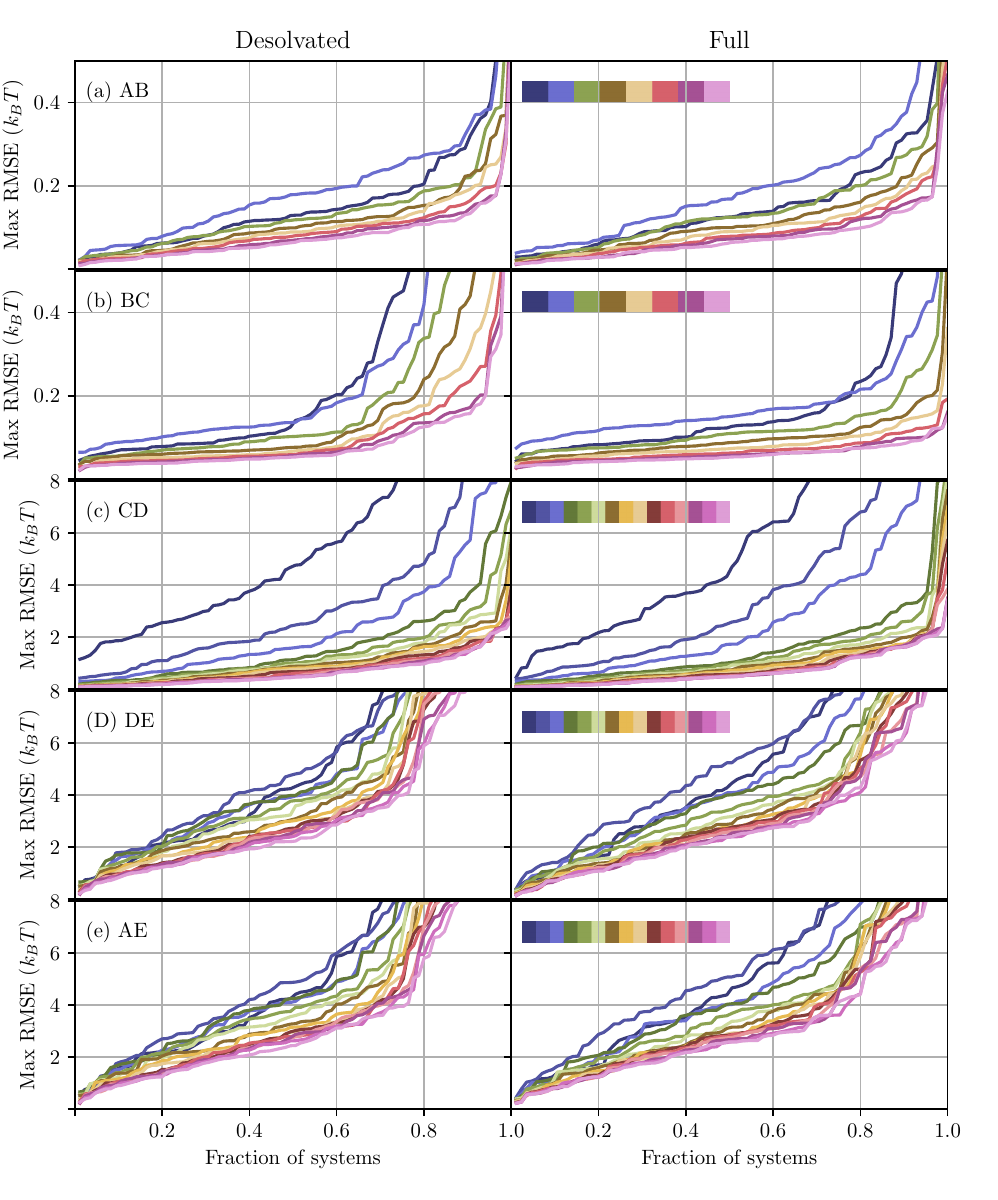}
\end{center}
\caption{
{\bf Fraction of systems} with free energy differences calculated within a certain root mean square error of the final value. The rows are for free energy differences between different pairs of milestones. The columns are for the Desolvated (left) and Full (right) solvation options. Each line indicates a different number of cycles, with a total of 8 for the top two rows and 15 for the remainder. In the sequence of colors inset in the right column of each row, the color indicate an increasing number of cycles from left to right.
\label{fig:convergence}}
\end{figure}

With the exception of $f_{DE}$, free energy differences between pairs of adjacent milestones in the thermodynamic cycle are feasible to calculate with AlGDock, converging quickly or at a steady rate. To evaluate the convergence of these estimates as a function of simulation time, the root mean square error (RMSE) between estimates for a particular cycle and estimates after the final cycle (for a particular solvation option) is considered. In the final cycle, the RMSE is simply the standard deviation. The fraction of systems with RMSE less than certain values is reported for different numbers of cycles in Figure \ref{fig:convergence}.

The free energy difference between milestones A and B, associated with transferring the ligand between the sampling and target force fields, converges quickly. After the first cycle, the RMSE of $f_{AB}$ is less than 1 $k_B T$ for all systems. After eight cycles, it is less than 0.5 $k_B T$ for all but one set of calculations, 0.544 $k_B$T for 1p62 with the Desolvated option. The fast convergence of $f_{AB}$ is expected because milestones A and B are meant to be the same force field, but simply evaluated using different software packages. Convergence is only dependent on adequate sampling of milestone B. 

The free energy difference between milestones B and C, associated with changing the temperature of the ligand (and removing implicit solvent in the Desolvated option), also converges quickly. For the Desolvated option, the RMSE of $f_{BC}$ is less than 2 $k_B T$ for all but 1jje (3.2 $k_B T$) after the first cycle and less than 1 $k_B T$ after eight cycles. The Full option converges more quickly, with a RMSE within 1 $k_B T$ after one cycle and within 0.15 $k_B T$ after eight. The faster convergence of the Full option is reasonable because milestones B and C are both in implicit solvent, implying that the protocol involves moving across a smaller region of configuration space. 

The free energy difference between milestones C and D, associated with changing the temperature of the ligand and scaling the receptor-ligand interaction grid, converges more slowly than between milestones B and C. For the Desolvated option, the RMSE of $f_{CD}$ is over 8 $k_B T$ for 23 systems after one cycle. However, after 15 cycles, the RMSE is below 2.5 $k_B T$ for all but one system, 2.65 $k_B T$ for 1t40. For the Full option, the RMSE of $f_{CD}$ is over 8 $k_B T$ for 28 systems after one cycle. However, after 15 cycles, it is likewise below 2.5 $k_B T$ for all but one system, 3.41 $k_B T$ for 1l7f. For both solvation protocols, the RMSE of $f_{CD}$ is below 2 $k_B T$ for about 80\% of systems after five cycles (Figure \ref{fig:convergence}). It is reasonable that $f_{CD}$ converges more slowly than $f_{BC}$ because adding protein-ligand interactions leads to many local minima that simulations may become trapped in.

In contrast to the other pairs of milestones, the free energy difference between milestones D and E does not converge quickly nor at a steady rate for as many systems. For the Desolvated option, the RMSE is above 8 $k_B T$ for 26 systems after one cycle. After 15 cycles, it is still above 8 $k_B T$ for 9 systems. For the Full option, the RMSE is above 8 $k_B T$ for 23 systems after one cycle and 5 systems after 15 cycles. With both protocols, the RMSE is below 2.5 $k_B T$ for about 60\% of systems and 4 $k_B T$ for about 80\% of systems. After about 5 cycles, the speed at which precision improves substantially slows. Given minimal changes in RMSE after about 12 cycles, it is unlikely that extending simulation beyond 15 cycles would lead to significant reduction in RMSE.

The convergence of $f_{DE}$ appears to be the limiting factor in the convergence of the BPMF. If the convergence of the total BPMF were limited by different pairs of milestones depending on the system, then the RMSE curves in the final row of Figure \ref{fig:convergence} would differ from any other row. If one particular pair of milestones always limits the convergence of total BPMF, the curves on the bottom row would resemble curves for that pair of milestones. In the data, the RMSE curves in the penultimate and final row of Figure \ref{fig:convergence} are nearly identical. 

\section*{\sffamily \Large Full solvation is less susceptible to false convergence}

The prior discussion of standard deviations and RMSEs neglects the possibility of false convergence. False convergence occurs when thermodynamic expectations and free energy differences are apparently stable, even across multiple independent simulations, but simulations do not truly sample from the relevant configuration space. In general, false convergence is difficult to diagnose. However, in the context of AlGDock calculations, it is possible to compare results from the Desolvated and Full solvation options. For these options, although intermediate milestones differ, milestones A and E are equivalent and calculations should, in principle, lead to the same total BPMF. 

\begin{figure}[p]
\begin{center}
\includegraphics[width=3.18in]{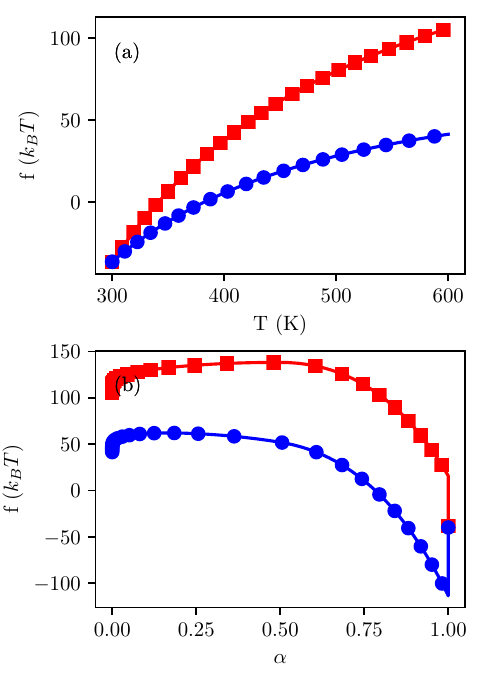}
\end{center}
\caption{
{\bf Representative cumulative free energies as a function of progress} for states (a) AC and (b) DE for Desolvated (red squares) or Full (blue circles) options. Simulations are of PDB ID 1gpk. Panel (a) is based on the sum of $f_{AB}$ and the free energy difference between $T_T$ = 300 K and the temperature on the x axis. Panel (b) is based on the sum of $f_{AC}$ and the free energy difference between $\alpha=0$ and the progress variable on the x axis. The final point on the plot shows $f_{AE}$ for both pathways. For clarity, markers are shown only every four thermodynamic states. 
\label{fig:pathways}}
\end{figure}

Figure \ref{fig:pathways} shows free energies from a representative pair of simulations that take different pathways but yield nearly equivalent BPMFs. $f_{AB}$ is the same pair of states for both pathways. The magnitude of $f_{BC}$ is larger for the Desolvated pathway, consistent with the removal of implicit solvent in addition to the temperature change. $f_{CD}$ has a similar overall shape. The BPMFs match because $f_{DE}$ have opposite signs.

In the majority of systems (63 systems or 74.1\%), the mean BPMF from Desolvated and Full solvation options is within error (the sum of standard deviations for each of the estimates). In other cases, either the mean BPMF from Desolvated (5 systems) or Full (17 systems) solvation options is lower (Figure \ref{fig:false_convergence}). Most of these differences are less than 10 $k_B T$, but several are very large, up to around 100 $k_B T$. 

To facilitate analysis of false convergence, the consensus BPMF was defined as the lower of mean BPMFs for the two solvation options. The RMSE was computed relative to the consensus BPMF rather than the final BPMF observed for a specific solvation option. The fraction of systems with RMSE less than certain values is reported for different numbers of cycles in Figure S3 of the Supplementary Material. Reflecting its higher propensity for false convergence, the curves for the Desolvated option are noticeably shifted up and to the left compare to the RMSE versus the final value. After the last cycle, many systems still have high RMSEs relative to the consensus: 35.2\% $>$ 5 $k_B T$, 15.2\% $>$ 8 $k_B T$, 11.7\% $>$ 20 $k_B T$, and 9.4\% $>$ 30 $k_B T$. Curves for the Full option are shifted upwards more subtly. After the last cycle, fewer systems still have high RMSEs relative to the consensus: 35.2\% $>$ 4 $k_B T$, 24.7\% $>$ 5 $k_B T$, and 9.4\% $>$ 8 $k_B T$. The largest RMSE is 12.8 $k_B T$.

\section*{\sffamily \Large Convergence is limited by differences in sampled configuration space}

The false convergence observed in some simulations is caused by differences in the important configuration space of milestones D and E.
In general, the convergence of calculated free energy differences between a pair of thermodynamic states is facilitated by overlap of the configuration space important to the states \cite{Wood1991}. The described $f_{DE}$ calculations are based on samples from milestone D. If the configuration space important to milestones D and E differ, then sampling configurations important to milestone E using simulations of milestone D is a rare event. 

\begin{figure}[p]
\begin{center}
\includegraphics[width=3.18in]{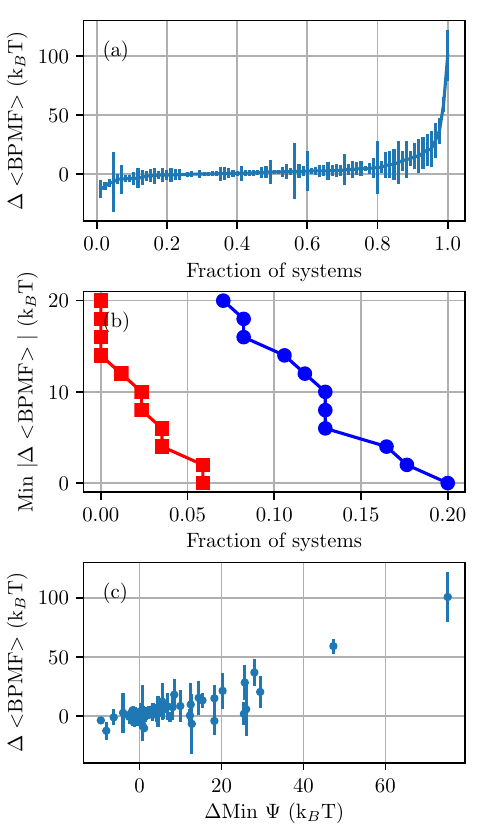}
\end{center}
\caption{
{\bf Differences between the mean BPMF of Desolvated and Full solvation options.} (a) Data points are ordered by the difference in mean BPMFs and the error bars are the sum of the standard deviations of the two estimates. (b) The fraction of systems in which the difference in mean BPMFs is larger than the sum of standard deviations of the two estimates. Either the Desolvated (red squares) or Full (blue circles) BPMF is lower by at least a certain value. (c) The difference in the minimum interaction energy, according to the force field in milestone E, versus the difference in mean BPMFs.
\label{fig:false_convergence}}
\end{figure}

Although it is difficult to determine whether the important configuration space of milestone E has been adequately sampled, it is feasible to compare the configuration space assessed with different solvation options. If they do access the same space, then it is reasonable to expect $f_{DE}$ and BPMF estimates to be consistent. If they do not, then $f_{DE}$ and thereby BPMFs are likely to be different. One way to quantify whether simulations access the configuration space important to milestone E is the minimum interaction energy, according to the force field used in milestone E. In a significant subset of systems, the minimum interaction energy observed in samples from milestone D with the two solvation options substantially differs (Figure \ref{fig:false_convergence}c). Indeed, large differences the minimum interaction energy appear to be a necessary, but not sufficient, condition for large differences in BPMFs between the Desolvated and Full solvation options. That is, all cases with large differences in BPMFs also have large differences in the minimum interaction energy. However, there are a few cases in which large differences in the minimum interaction energy do not correspond to large differences in the mean BPMF.

While it may be intuitive to think that large differences in the mean interaction energy are due to large differences in binding poses, the data suggest that this is not always the case. In a number of simulations, a large difference in interaction energy (on the order of 100 $k_BT$) is associated with an RMSD of less that 1 \AA~from the crystallographic and the minimum-energy configuration (Figure S4 of the Supplementary Material). Furthermore, there are simulations in which configuration with the lowest RMSD compared to the native pose is greater than 2 \AA, but the difference in the minimum interaction energy is minimal. In these cases, a number of distinct conformations may have a similar interaction energy.

In cases where the two solvation options differ in sampled configurations and interaction energies, one option is not adequately sampling the configuration space in milestone E. More precise calculations of $f_{DE}$ would require either introducing intermediate thermodynamic states, adding to the computational cost, or using a different force field at milestone D that it is more similar to milestone E.

\section*{\sffamily \Large BPMF calculations and interaction energies are similarly successful in identifying native poses}

The most common strategy for ranking a set of binding poses to a receptor is based on the interaction energy. It is also reasonable to consider the total energy of the complex, in which the internal energy of the ligand also differs between poses. In principle, a better ranking scheme would account for the relative entropy of each pose and be based on the free energy of the pose. For this reason, I considered whether BPMF calculations, which account for the configurational entropy of the ligand but not the receptor, outperform strategies based on individual configurations (interaction and total energy) at ranking binding poses. For different sets of structures, the scoring functions (interaction energy, total energy, or pose-specific BPMF) were evaluated based on whether a native pose (RMSD $<$ 2 \AA~ from the crystal) has the lowest or close-to-lowest score (Table \ref{tab:fraction_native_energy}).

If the crystallographic pose and predicted poses from UCSF DOCK 6 are considered, the minimum interaction energy strategy is fairly successful. The pose with the lowest interaction energy is native in 81.2\% of systems according to the UCSF DOCK 6 grid score in 87.1\% of systems according to the interaction energy based from milestone E. The lowest internal energy is an even better strategy, correctly identifying the native pose in 92.9\% of systems. (Due to decoys observed during simulation, this success rate is reduced to 88.2\% when considering poses from all the BPMF simulations.) In 95.3\% of systems, the native pose has an interaction energy or total energy no more than 8 $k_BT$ higher the lowest-energy pose. Several reasons for the high success rate of these methods is because self-docking opposed to cross-docking was performed, only poses with a center of mass within 6 \AA~ of the crystallographic binding pose were considered, and because of the extensive sampling of binding poses. The likely reason that milestone E outperforms the UCSF DOCK 6 grid score is because it incorporates solvation free energies.

Although all BPMF calculations included native poses among the initial structures, some simulations drifted away from native conformations over the course of equilibration. This issue was more prevalent with the Desolvated than the Full solvation option; the native pose was observed during production of milestone D in 90.4\% of simulations with the former and 94.9\% of simulations with the latter solvation option. Due to this configuration space drift, the minimum interaction energy and total energy were less successful at identifying the native pose based on samples from BPMF calculations than from initial starting structures (which all included native poses).

When considering individual BPMF calculations, the success of native pose identification was more dependent on the force field than whether scores were based on interaction energies, total energies, or pose-specific BPMFs. For ranking observed poses, the force field at milestone E was best, milestone D with Full solvation was second, and milestone D with the Desolvated option was the worst among tested force fields. On the other hand, the interaction energy, total energy, and pose-specific BPMF performed similarly. 

\begin{figure}[p]
\begin{center}
\includegraphics[width=3.18in]{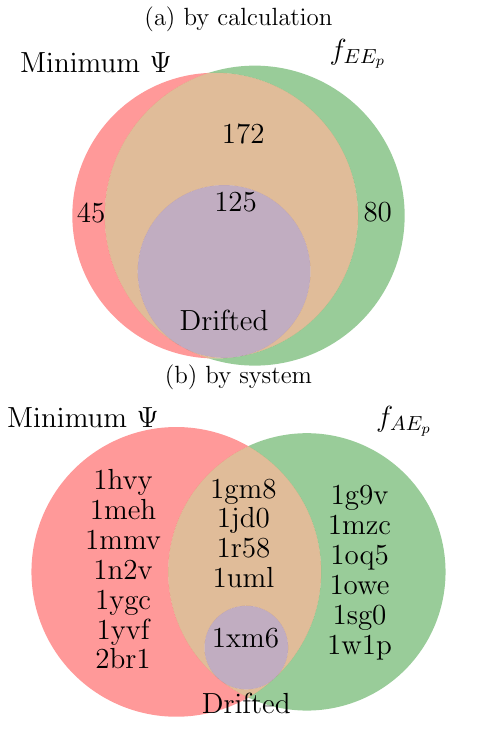}
\end{center}
\caption{
{\bf Venn diagram of geometric decoys.} Incorrect binding pose predictions, or geometric decoys, arise when a nonnative pose has a lower score than any native pose. The Venn diagrams show the overlap between sets of (a) BPMF calculations or (b) systems in which there are geometric decoys according to the interaction energy in milestone E (Minimum $\Psi$), the pose-specific BPMF ($f_{EE_p}$ in panel (a) and $f_{AE_p}$ in panel (b)), or because no native poses were observed (Drifted). Panel (a) is labeled by the number of BPMF calculations and is based on poses observed in each calculation. Panel (b) is labeled by PDB identifiers for the particular systems and is based on poses observed in all BPMF calculations for a system.
\label{fig:decoys}}
\end{figure}

Even though overall performance of native pose identification was similar among scoring schemes, the schemes often failed in different BPMF calculations and systems (Figure \ref{fig:decoys}). In the majority of BPMF calculations where the binding pose was incorrectly identified, both the minimum interaction energy and pose-specific BPMF were deceived by geometric decoys (172) or did not sample the native state (125). A slightly larger number of calculations incorrectly identified the binding pose based on the pose-specific BPMF exclusively (80) than based on the minimum interaction energy exclusively (45). When considering \emph{all} BPMF calculations for a specific system, the pose-specific BPMF was comparably reliable with the interaction energy. Whereas the pose-specific BPMF exclusively misidentified the native pose in 6 systems, the minimum interaction energy was exclusively incorrect in 7 systems.

\section*{\sffamily \Large DISCUSSION}

\section*{\sffamily \Large Thermodynamic state initialization and automatic adaptation reliably yields reasonable protocols}

The selection of intermediate states between two thermodynamic milestones of interest is a ubiquitous problem in molecular simulation. The usual approach to this problem is an iterative trial-and-error process starting from a naive protocol, checking for issues such as replica exchange bottlenecks, and manually inserting and removing states as necessary. I have developed a simple and robust approach to initialize a series of thermodynamic states based on only a single adjustable parameter, the thermodynamic speed. In the vast majority of cases, the initialization protocol yielded consistent replica exchange rates across neighboring thermodynamic states without further fine-tuning. I also developed an automated approach to add additional states when the observed replica exchange rate falls below 40\%. With this procedure, I was able to run a large number of simulations on a diverse array of protein-ligand complexes without manual intervention. The described approach to trailblazing and adapting thermodynamic state space may find use in other classes of simulations.

Replica exchange calculations in this present study include more thermodynamic states than most published molecular simulations. Conventional wisdom about replica exchange is that an optimal number of replicas will maximize efficiency. With too few replicas, there is limited configuration space overlap between neighbors and exchange rates are vanishingly small. With too many replicas, metrics of replica exchange efficiency, such as a mean round-trip time, diminish. However, in a recent study involving extensive simulation of several distinct processes, my group found that if there are no bottlenecks in which the replica exchange rate is below 40\%, the number of states has little impact on the convergence of free energy estimates \cite{Nguyen2016}. Hence, I chose to include a large number of states to minimize the probability that later sampling will explore different regions of configuration space and reduce the exchange rate between neighboring states.

While useful, the described thermodynamic state initialization process remains imperfect. Replica exchange is particularly beneficial when the important configuration space of a thermodynamic state is a subset of the important configuration space of another. The present procedure is limited to the variation of a single thermodynamic parameter between milestones. Future improvements could accommodate varying multiple parameters (e.g. separate parameters for the temperature, van der Waals grids, and electrostatic grids) in between thermodynamic states of interest to promote sampling while minimizing unnecessary traversals of configuration space.

\section*{\sffamily \Large Most free energy differences between sampled states are converged}

It has been argued that many simulations of biomolecular binding processes do not adequately sample relevant configuration space and therefore results are not converged and are irreproducible \cite{Mobley2012b}. Within his discussion of problematic degrees of freedom, \citet{Mobley2012b} highlighted ligand binding modes and internal conformational changes in small molecules as common reasons for failed convergence. In the majority of simulations considered in this paper, $f_{AD}$ estimates are very precise. Achieving precise free energy differences between milestones A and D requires adequate sampling of both of the aforementioned problematic classes of conformational transitions. The precise results suggest that the described approach is successful at addressing the sampling problems for \emph{simulated} thermodynamic states.

\section*{\sffamily \Large BPMF convergence and native pose prediction is predicated on overlap between milestones D and E}

While free energy differences between milestones A and D were precise in nearly all systems, the precision of $f_{DE}$ estimates was more variable, limiting the convergence of BPMF calculations and the accuracy of native pose prediction. The performance of $f_{DE}$ estimates and native pose prediction was highly dependent on whether the ligand was considered desolvated or fully solvated when bound to the protein, implicating the solvation option as a key factor in the overlap between the important configuration spaces of milestones D and E. The strong performance of the interaction energy based on milestone E in identifying the native binding pose suggests that it is desirable to bring sampling closer to the importance configuration space of milestone E (opposed to making milestone E more like milestone D with either solvation option). In future work, more precise BPMF estimates may be attained by introducing intermediate states or by altering the force field for milestone D to become more similar to milestone E. For example, milestone D could be based on a grid that uses generalized Born instead of Poisson-Boltzmann electrostatics. A related possibility is to model desolvation of the ligand due to the receptor using a grid-based fractional desolvation term \cite{Mysinger2010}. The strong performance of the consensus binding pose prediction based on \emph{all} BPMF estimates for a system suggests that more precise BPMF calculations will also yield improved binding pose prediction.

Even with shortcomings in BPMF convergence, the present native pose prediction strategies perform comparably to other docking programs (Table \ref{tab:fraction_native_energy}). Because the Astex Diverse Set \cite{Hartshorn2007} is a widely used benchmark, I will only mention a few results. In the original paper on the dataset, the standard GOLD protocol predicted the native pose within 2 \AA~ in 80.5\% of systems \cite{Hartshorn2007}. With best-practice structures, GLIDE was successful according to the same criterion for in 82\% of systems (Figure 1 of \citet{Repasky2012}). ICM has particularly strong performance, successful in 91\% of systems \cite{Neves2012}. For comparison, when ranking poses with the free energy based on the force field in milestone E, the calculations described herein are successful in 75.8\% of Desolvated and 83.9\% of Full BPMF calculations (Table \ref{tab:fraction_native_energy}). A caveat to this comparison is that the BPMF calculations were started with the native pose (as well as poses generated by UCSF DOCK 6), but other methods were required to sample the binding pose \emph{de novo}. Another helpful reference point is a recent comprehensive evaluation by \citet{Wang2016d}, who found that 10 programs were successful in 40\% to 60\% of complexes in the PDBbind refined set (version 2014).

\section*{\sffamily \Large Ligand electrostatics in the protein environment are better treated as fully solvated}

One of the more surprising results is that the Full solvation option outperforms the Desolvated option by yielding lower BPMFs and better binding pose predictions. After all, a ligand that is bound to a protein must shed most if not all of its hydration shell! A possible explanation is that many protein binding sites could mimic the dielectric environment of water. To facilitate protein folding and solubility, soluble proteins usually contain hydrophobic residues in the interior and hydrophilic residues on the exterior. Since binding sites are primarily on protein surfaces, bound ligands may be surrounded by residues whose dielectric behavior resembles water.

\section*{\sffamily \Large CONCLUSIONS}

I have developed a reasonably robust method to estimate BPMFs for protein-ligand systems. The largest sources of imprecision are found to be configuration space overlap between representations of the complex. 

\section*{\sffamily \Large ACKNOWLEDGMENTS}


I thank Michael Shirts (University of Colorado Boulder), John Chodera (MSKCC), and Trung Hai Nguyen (IIT) for helpful discussions. Peter Eastman (Simbios) and John Chodera assisted with implementing an early version of the code in OpenMM. Michael Sherman (Simbios) made an invaluable suggestion of using precomputed grids.

Versions of this software were tested using different computing resources. I think Rob Gardner, Lincoln Bryant, and Balamurugan Desinghu (Open Science Grid) and Tom Milledge (Duke Shared Computing Resources) for assistance with using their resources. Most reported calculations were performed on the Open Science Grid \cite{Pordes2007, Sfiligoi2009}, which is supported by the National Science Foundation and the U.S. Department of Energy's Office of Science. Benchmark calculations were performed with the Extreme Science and Engineering Discovery Environment (XSEDE) \citep{xsede}, which is supported by National Science Foundation grant number ACI-1548562. Use of XSEDE Comet at the San Diego Supercomputer Center was provided through allocation TG-MCB150144. Test calculations were performed using Duke Shared Computing Resources and my research group's cluster at IIT.

I thank OpenEye Scientific Software, Inc. and UCSF for providing academic licenses to their software.

In the early stages of this project, David Beratan (Duke) was a supportive postdoctoral advisor.
At the beginning of this project, I was a postdoctoral scholar at Duke, supported by NIH 2P50 GM067082-06-10. Later in the project, I was supported by NIH 1R15 GM114781. I also spent a month at Stanford with Simbios as an OpenMM visiting scholar.

\clearpage




\clearpage

\section{Tables}

\begin{table}[ht]
\begin{tabular}{| l | c | c | c | c | c | c |}  
\hline
Samples & \multicolumn{2}{c}{Scoring}  & \multicolumn{4}{|c|}{Native within energy of minimum ($k_BT$)} \\
\hline
 & Force Field & Score Type & 8 & 4 & 2 & 0 \\
\hline
 \multirow{3}{*}{\shortstack[1]{xtal \\ + DOCK 6}} 
 & D6 & min $\Psi$ & 0.941 (0.026) & 0.906 (0.032) & 0.847 (0.039) & 0.812 (0.042) \\
 & E & min u & 0.953 (0.023) & 0.953 (0.023) & 0.929 (0.028) & 0.929 (0.028) \\
 & E & min $\Psi$ & 0.953 (0.023) & 0.918 (0.030) & 0.882 (0.035) & 0.871 (0.036) \\
 \hline
 \multirow{12}{*}{Desolvated} 
 & D & min u & 0.845 (0.012) & 0.781 (0.014) & 0.732 (0.014) & 0.682 (0.015) \\
 & D & mean u & 0.847 (0.012) & 0.769 (0.014) & 0.705 (0.015) & 0.653 (0.016) \\
 & D & fe u & 0.919 (0.009) & 0.912 (0.009) & 0.873 (0.011) & 0.690 (0.015) \\
 & D & min Psi & 0.886 (0.010) & 0.826 (0.012) & 0.786 (0.013) & 0.716 (0.015) \\
 & D & mean Psi & 0.890 (0.010) & 0.842 (0.012) & 0.803 (0.013) & 0.714 (0.015) \\
 & D & fe Psi & 0.919 (0.009) & 0.912 (0.009) & 0.873 (0.011) & 0.690 (0.015) \\
 & E & min u & 0.833 (0.012) & 0.811 (0.013) & 0.788 (0.013) & 0.771 (0.014) \\
 & E & mean u & 0.830 (0.012) & 0.796 (0.013) & 0.772 (0.014) & 0.752 (0.014) \\
 & E & fe u & 0.842 (0.012) & 0.813 (0.013) & 0.787 (0.013) & 0.758 (0.014) \\
 & E & min Psi & 0.848 (0.012) & 0.831 (0.012) & 0.814 (0.013) & 0.788 (0.013) \\
 & E & mean Psi & 0.842 (0.012) & 0.818 (0.013) & 0.788 (0.013) & 0.757 (0.014) \\
 & E & fe Psi & 0.856 (0.011) & 0.818 (0.013) & 0.799 (0.013) & 0.767 (0.014) \\
\hline
 \multirow{12}{*}{Full}
 & D & min u & 0.883 (0.010) & 0.804 (0.013) & 0.753 (0.014) & 0.699 (0.015) \\
 & D & mean u & 0.905 (0.010) & 0.825 (0.012) & 0.759 (0.014) & 0.682 (0.015) \\
 & D & fe u & 0.948 (0.007) & 0.943 (0.008) & 0.899 (0.010) & 0.702 (0.015) \\
 & D & min Psi & 0.910 (0.009) & 0.867 (0.011) & 0.816 (0.013) & 0.741 (0.014) \\
 & D & mean Psi & 0.921 (0.009) & 0.872 (0.011) & 0.831 (0.012) & 0.734 (0.014) \\
 & D & fe Psi & 0.948 (0.007) & 0.943 (0.008) & 0.899 (0.010) & 0.702 (0.015) \\
 & E & min u & 0.897 (0.010) & 0.866 (0.011) & 0.843 (0.012) & 0.820 (0.013) \\
 & E & mean u & 0.893 (0.010) & 0.863 (0.011) & 0.839 (0.012) & 0.807 (0.013) \\
 & E & fe u & 0.908 (0.009) & 0.879 (0.011) & 0.864 (0.011) & 0.839 (0.012) \\
 & E & min Psi & 0.910 (0.009) & 0.893 (0.010) & 0.876 (0.011) & 0.846 (0.012) \\
 & E & mean Psi & 0.897 (0.010) & 0.880 (0.011) & 0.864 (0.011) & 0.834 (0.012) \\
 & E & fe Psi & 0.913 (0.009) & 0.884 (0.010) & 0.868 (0.011) & 0.840 (0.012) \\
 \hline
\multirow{2}{*}{All} 
 & E & min u & 0.918 (0.030) & 0.906 (0.032) & 0.882 (0.035) & 0.882 (0.035) \\
 & E & min $\Psi$ & 0.941 (0.026) & 0.918 (0.030) & 0.894 (0.033) & 0.871 (0.036) \\
 \hline
    \end{tabular}  
  \caption{Fraction of calculations in which a native binding pose (RMSD from xtal structure $<$ 2 \AA) is within a specified cutoff of the minimum-energy structure. Binding poses were obtained either by scoring the crystal structure and docking (xtal + DOCK 6), from BPMF calculations with the Desolvated or Full options at milestone D, or all of the above. Docking poses were the 50 best-scoring poses from docking in which the ligand center of mass is within 6 \AA~of the center of mass of the crystallographic pose. Scoring was based on one of three force fields: the UCSF DOCK 6 grid score (D6), milestone D, or milestone E. The score was either the minimum or mean interaction energy $\Psi$, minimum or mean total energy $u$, or free energy based on reweighing the interaction energy (Equation \ref{eq:reweighing_interact} and \ref{eq:pose_free_energy}) or total energy (Equation \ref{eq:reweighing_all} and \ref{eq:pose_free_energy}). Parentheses contain the standard error of the sample proportion, $\sqrt{p(1-p)/n}$.
\label{tab:fraction_native_energy}}
\end{table}


\clearpage


\begin{thebibliography}{49}
\expandafter\ifx\csname natexlab\endcsname\relax\def\natexlab#1{#1}\fi
\expandafter\ifx\csname bibnamefont\endcsname\relax
  \def\bibnamefont#1{#1}\fi
\expandafter\ifx\csname bibfnamefont\endcsname\relax
  \def\bibfnamefont#1{#1}\fi
\expandafter\ifx\csname citenamefont\endcsname\relax
  \def\citenamefont#1{#1}\fi
\expandafter\ifx\csname url\endcsname\relax
  \def\url#1{\texttt{#1}}\fi
\expandafter\ifx\csname urlprefix\endcsname\relax\def\urlprefix{URL }\fi
\providecommand{\bibinfo}[2]{#2}
\providecommand{\eprint}[2][]{\url{#2}}

\bibitem[{\citenamefont{Minh}(2012)}]{Minh2012}
\bibinfo{author}{\bibfnamefont{D.~D.~L.} \bibnamefont{Minh}},
  \bibinfo{journal}{J. Chem. Phys.} \textbf{\bibinfo{volume}{137}},
  \bibinfo{pages}{104106} (\bibinfo{year}{2012}).

\bibitem[{\citenamefont{Nguyen and Minh}(2018)}]{Nguyen2018}
\bibinfo{author}{\bibfnamefont{T.~H.} \bibnamefont{Nguyen}} \bibnamefont{and}
  \bibinfo{author}{\bibfnamefont{D.~D.~L.} \bibnamefont{Minh}},
  \bibinfo{journal}{J. Chem. Phys.} \textbf{\bibinfo{volume}{148}},
  \bibinfo{pages}{104114} (\bibinfo{year}{2018}).

\bibitem[{\citenamefont{Mobley et~al.}(2007)\citenamefont{Mobley, Graves,
  Chodera, McReynolds, Shoichet, and Dill}}]{Mobley2007}
\bibinfo{author}{\bibfnamefont{D.~L.} \bibnamefont{Mobley}},
  \bibinfo{author}{\bibfnamefont{A.~P.} \bibnamefont{Graves}},
  \bibinfo{author}{\bibfnamefont{J.~D.} \bibnamefont{Chodera}},
  \bibinfo{author}{\bibfnamefont{A.~C.} \bibnamefont{McReynolds}},
  \bibinfo{author}{\bibfnamefont{B.~K.} \bibnamefont{Shoichet}},
  \bibnamefont{and} \bibinfo{author}{\bibfnamefont{K.~A.} \bibnamefont{Dill}},
  \bibinfo{journal}{J. Mol. Biol.} \textbf{\bibinfo{volume}{371}},
  \bibinfo{pages}{1118} (\bibinfo{year}{2007}).

\bibitem[{\citenamefont{Ucisik et~al.}(2014)\citenamefont{Ucisik, Zheng, Faver,
  and Merz}}]{Ucisik2014}
\bibinfo{author}{\bibfnamefont{M.~N.} \bibnamefont{Ucisik}},
  \bibinfo{author}{\bibfnamefont{Z.}~\bibnamefont{Zheng}},
  \bibinfo{author}{\bibfnamefont{J.~C.} \bibnamefont{Faver}}, \bibnamefont{and}
  \bibinfo{author}{\bibfnamefont{K.~M.} \bibnamefont{Merz}},
  \bibinfo{journal}{J. Chem. Theory Comput.} \textbf{\bibinfo{volume}{10}},
  \bibinfo{pages}{1314} (\bibinfo{year}{2014}).

\bibitem[{\citenamefont{Xie et~al.}(2017)\citenamefont{Xie, Nguyen, and
  Minh}}]{Xie2017}
\bibinfo{author}{\bibfnamefont{B.}~\bibnamefont{Xie}},
  \bibinfo{author}{\bibfnamefont{T.~H.} \bibnamefont{Nguyen}},
  \bibnamefont{and} \bibinfo{author}{\bibfnamefont{D.~D.~L.}
  \bibnamefont{Minh}}, \bibinfo{journal}{J. Chem. Theory Comput.}
  \textbf{\bibinfo{volume}{13}}, \bibinfo{pages}{2930} (\bibinfo{year}{2017}).

\bibitem[{\citenamefont{Nguyen et~al.}(2018)\citenamefont{Nguyen, Zhou, and
  Minh}}]{Nguyen2018a}
\bibinfo{author}{\bibfnamefont{T.~H.} \bibnamefont{Nguyen}},
  \bibinfo{author}{\bibfnamefont{H.-X.} \bibnamefont{Zhou}}, \bibnamefont{and}
  \bibinfo{author}{\bibfnamefont{D.~D.~L.} \bibnamefont{Minh}},
  \bibinfo{journal}{J. Comput. Chem.} \textbf{\bibinfo{volume}{39}},
  \bibinfo{pages}{621} (\bibinfo{year}{2018}).

\bibitem[{\citenamefont{Minh}(2015)}]{Minh2015}
\bibinfo{author}{\bibfnamefont{D.~D.~L.} \bibnamefont{Minh}},
  \bibinfo{journal}{arXiv} \bibinfo{pages}{1507.03703v1}
  (\bibinfo{year}{2015}).

\bibitem[{\citenamefont{Xie and Minh}(2019)}]{Xie2019}
\bibinfo{author}{\bibfnamefont{B.}~\bibnamefont{Xie}} \bibnamefont{and}
  \bibinfo{author}{\bibfnamefont{D.~D.~L.} \bibnamefont{Minh}},
  \bibinfo{journal}{J. Comput.-Aided Mol. Des.} \textbf{\bibinfo{volume}{33}},
  \bibinfo{pages}{61} (\bibinfo{year}{2019}).

\bibitem[{\citenamefont{Hartshorn et~al.}(2007)\citenamefont{Hartshorn,
  Verdonk, Chessari, Brewerton, Mooij, Mortenson, and Murray}}]{Hartshorn2007}
\bibinfo{author}{\bibfnamefont{M.~J.} \bibnamefont{Hartshorn}},
  \bibinfo{author}{\bibfnamefont{M.~L.} \bibnamefont{Verdonk}},
  \bibinfo{author}{\bibfnamefont{G.}~\bibnamefont{Chessari}},
  \bibinfo{author}{\bibfnamefont{S.~C.} \bibnamefont{Brewerton}},
  \bibinfo{author}{\bibfnamefont{W.~T.~M.} \bibnamefont{Mooij}},
  \bibinfo{author}{\bibfnamefont{P.~N.} \bibnamefont{Mortenson}},
  \bibnamefont{and} \bibinfo{author}{\bibfnamefont{C.~W.}
  \bibnamefont{Murray}}, \bibinfo{journal}{J. Med. Chem.}
  \textbf{\bibinfo{volume}{50}}, \bibinfo{pages}{726} (\bibinfo{year}{2007}).

\bibitem[{\citenamefont{Gallicchio and Levy}(2012)}]{Gallicchio2012}
\bibinfo{author}{\bibfnamefont{E.}~\bibnamefont{Gallicchio}} \bibnamefont{and}
  \bibinfo{author}{\bibfnamefont{R.~M.} \bibnamefont{Levy}},
  \bibinfo{journal}{J. Comput.-Aided Mol. Des.} \textbf{\bibinfo{volume}{26}},
  \bibinfo{pages}{505} (\bibinfo{year}{2012}).

\bibitem[{\citenamefont{Wang et~al.}(2013)\citenamefont{Wang, Chodera, Yang,
  and Shirts}}]{Wang2013b}
\bibinfo{author}{\bibfnamefont{K.}~\bibnamefont{Wang}},
  \bibinfo{author}{\bibfnamefont{J.~D.} \bibnamefont{Chodera}},
  \bibinfo{author}{\bibfnamefont{Y.}~\bibnamefont{Yang}}, \bibnamefont{and}
  \bibinfo{author}{\bibfnamefont{M.~R.} \bibnamefont{Shirts}},
  \bibinfo{journal}{J. Comput.-Aided Mol. Des.} \textbf{\bibinfo{volume}{27}},
  \bibinfo{pages}{989} (\bibinfo{year}{2013}).

\bibitem[{\citenamefont{Wang et~al.}(2015)\citenamefont{Wang, Wu, Deng, Kim,
  Pierce, Krilov, Lupyan, Robinson, Dahlgren, Greenwood et~al.}}]{Wang2015}
\bibinfo{author}{\bibfnamefont{L.}~\bibnamefont{Wang}},
  \bibinfo{author}{\bibfnamefont{Y.}~\bibnamefont{Wu}},
  \bibinfo{author}{\bibfnamefont{Y.}~\bibnamefont{Deng}},
  \bibinfo{author}{\bibfnamefont{B.}~\bibnamefont{Kim}},
  \bibinfo{author}{\bibfnamefont{L.}~\bibnamefont{Pierce}},
  \bibinfo{author}{\bibfnamefont{G.}~\bibnamefont{Krilov}},
  \bibinfo{author}{\bibfnamefont{D.}~\bibnamefont{Lupyan}},
  \bibinfo{author}{\bibfnamefont{S.}~\bibnamefont{Robinson}},
  \bibinfo{author}{\bibfnamefont{M.~K.} \bibnamefont{Dahlgren}},
  \bibinfo{author}{\bibfnamefont{J.}~\bibnamefont{Greenwood}},
  \bibnamefont{et~al.}, \bibinfo{journal}{J. Am. Chem. Soc.}
  \textbf{\bibinfo{volume}{137}}, \bibinfo{pages}{2695} (\bibinfo{year}{2015}).

\bibitem[{\citenamefont{Pattabiraman et~al.}(1985)\citenamefont{Pattabiraman,
  Levitt, Ferrin, and Langridge}}]{Pattabiraman1985}
\bibinfo{author}{\bibfnamefont{N.}~\bibnamefont{Pattabiraman}},
  \bibinfo{author}{\bibfnamefont{M.}~\bibnamefont{Levitt}},
  \bibinfo{author}{\bibfnamefont{T.~E.} \bibnamefont{Ferrin}},
  \bibnamefont{and}
  \bibinfo{author}{\bibfnamefont{R.}~\bibnamefont{Langridge}},
  \bibinfo{journal}{J. Comput. Chem.} \textbf{\bibinfo{volume}{6}},
  \bibinfo{pages}{432} (\bibinfo{year}{1985}).

\bibitem[{\citenamefont{Meng et~al.}(1992)\citenamefont{Meng, Shoichet, and
  Kuntz}}]{Meng1992}
\bibinfo{author}{\bibfnamefont{E.~C.} \bibnamefont{Meng}},
  \bibinfo{author}{\bibfnamefont{B.~K.} \bibnamefont{Shoichet}},
  \bibnamefont{and} \bibinfo{author}{\bibfnamefont{I.~D.} \bibnamefont{Kuntz}},
  \bibinfo{journal}{J. Comput. Chem.} \textbf{\bibinfo{volume}{13}},
  \bibinfo{pages}{505} (\bibinfo{year}{1992}).

\bibitem[{\citenamefont{Minh}(2018)}]{Minh2018}
\bibinfo{author}{\bibfnamefont{D.~D.~L.} \bibnamefont{Minh}},
  \bibinfo{journal}{J. Comput. Chem.} \textbf{\bibinfo{volume}{39}},
  \bibinfo{pages}{1200} (\bibinfo{year}{2018}).

\bibitem[{\citenamefont{Hinsen}(2000)}]{Hinsen2000}
\bibinfo{author}{\bibfnamefont{K.}~\bibnamefont{Hinsen}}, \bibinfo{journal}{J.
  Comput. Chem.} \textbf{\bibinfo{volume}{21}}, \bibinfo{pages}{79}
  (\bibinfo{year}{2000}).

\bibitem[{\citenamefont{Shirts and Chodera}(2008)}]{Shirts2008}
\bibinfo{author}{\bibfnamefont{M.~R.} \bibnamefont{Shirts}} \bibnamefont{and}
  \bibinfo{author}{\bibfnamefont{J.~D.} \bibnamefont{Chodera}},
  \bibinfo{journal}{J. Chem. Phys.} \textbf{\bibinfo{volume}{129}},
  \bibinfo{pages}{124105} (\bibinfo{year}{2008}).

\bibitem[{\citenamefont{Wang et~al.}(2004)\citenamefont{Wang, Wolf, Caldwell,
  Kollman, and Case}}]{Wang2004a}
\bibinfo{author}{\bibfnamefont{J.}~\bibnamefont{Wang}},
  \bibinfo{author}{\bibfnamefont{R.~M.} \bibnamefont{Wolf}},
  \bibinfo{author}{\bibfnamefont{J.~W.} \bibnamefont{Caldwell}},
  \bibinfo{author}{\bibfnamefont{P.~A.} \bibnamefont{Kollman}},
  \bibnamefont{and} \bibinfo{author}{\bibfnamefont{D.~A.} \bibnamefont{Case}},
  \bibinfo{journal}{J. Comput. Chem.} \textbf{\bibinfo{volume}{25}},
  \bibinfo{pages}{1157} (\bibinfo{year}{2004}).

\bibitem[{\citenamefont{Jakalian et~al.}(1999)\citenamefont{Jakalian, Bush,
  Jack, and Bayly}}]{Jakalian1999}
\bibinfo{author}{\bibfnamefont{A.}~\bibnamefont{Jakalian}},
  \bibinfo{author}{\bibfnamefont{B.~L.} \bibnamefont{Bush}},
  \bibinfo{author}{\bibfnamefont{D.~B.} \bibnamefont{Jack}}, \bibnamefont{and}
  \bibinfo{author}{\bibfnamefont{C.~I.} \bibnamefont{Bayly}},
  \bibinfo{journal}{J. Comput. Chem.} \textbf{\bibinfo{volume}{21}},
  \bibinfo{pages}{132} (\bibinfo{year}{1999}).

\bibitem[{\citenamefont{Jakalian et~al.}(2002)\citenamefont{Jakalian, Jack, and
  Bayly}}]{Jakalian2002}
\bibinfo{author}{\bibfnamefont{A.}~\bibnamefont{Jakalian}},
  \bibinfo{author}{\bibfnamefont{D.~B.} \bibnamefont{Jack}}, \bibnamefont{and}
  \bibinfo{author}{\bibfnamefont{C.~I.} \bibnamefont{Bayly}},
  \bibinfo{journal}{J. Comput. Chem.} \textbf{\bibinfo{volume}{23}},
  \bibinfo{pages}{1623} (\bibinfo{year}{2002}).

\bibitem[{\citenamefont{Eastman and Pande}(2010)}]{Eastman2010}
\bibinfo{author}{\bibfnamefont{P.}~\bibnamefont{Eastman}} \bibnamefont{and}
  \bibinfo{author}{\bibfnamefont{V.~S.} \bibnamefont{Pande}},
  \bibinfo{journal}{Comput. Sci. Eng.} \textbf{\bibinfo{volume}{12}},
  \bibinfo{pages}{34} (\bibinfo{year}{2010}).

\bibitem[{\citenamefont{Onufriev et~al.}(2004)\citenamefont{Onufriev, Bashford,
  and Case}}]{Onufriev2004}
\bibinfo{author}{\bibfnamefont{A.}~\bibnamefont{Onufriev}},
  \bibinfo{author}{\bibfnamefont{D.}~\bibnamefont{Bashford}}, \bibnamefont{and}
  \bibinfo{author}{\bibfnamefont{D.~A.} \bibnamefont{Case}},
  \bibinfo{journal}{Proteins: Struct., Funct., Bioinf.}
  \textbf{\bibinfo{volume}{55}}, \bibinfo{pages}{383} (\bibinfo{year}{2004}).

\bibitem[{\citenamefont{Baker et~al.}(2001)\citenamefont{Baker, Sept, Joseph,
  Holst, and McCammon}}]{Baker2001}
\bibinfo{author}{\bibfnamefont{N.~A.} \bibnamefont{Baker}},
  \bibinfo{author}{\bibfnamefont{D.}~\bibnamefont{Sept}},
  \bibinfo{author}{\bibfnamefont{S.}~\bibnamefont{Joseph}},
  \bibinfo{author}{\bibfnamefont{M.~J.} \bibnamefont{Holst}}, \bibnamefont{and}
  \bibinfo{author}{\bibfnamefont{J.~A.} \bibnamefont{McCammon}},
  \bibinfo{journal}{Proc. Natl. Acad. Sci. USA} \textbf{\bibinfo{volume}{98}},
  \bibinfo{pages}{10037} (\bibinfo{year}{2001}).

\bibitem[{\citenamefont{Oberlin and Scheraga}(1998)}]{Oberlin1998}
\bibinfo{author}{\bibfnamefont{D.}~\bibnamefont{Oberlin}} \bibnamefont{and}
  \bibinfo{author}{\bibfnamefont{H.~A.} \bibnamefont{Scheraga}},
  \bibinfo{journal}{J. Comput. Chem.} \textbf{\bibinfo{volume}{19}},
  \bibinfo{pages}{71} (\bibinfo{year}{1998}).

\bibitem[{\citenamefont{Venkatachalam et~al.}(2003)\citenamefont{Venkatachalam,
  Jiang, Oldfield, and Waldman}}]{Venkatachalam2003}
\bibinfo{author}{\bibfnamefont{C.~M.} \bibnamefont{Venkatachalam}},
  \bibinfo{author}{\bibfnamefont{X.}~\bibnamefont{Jiang}},
  \bibinfo{author}{\bibfnamefont{T.}~\bibnamefont{Oldfield}}, \bibnamefont{and}
  \bibinfo{author}{\bibfnamefont{M.}~\bibnamefont{Waldman}},
  \bibinfo{journal}{J. Mol. Graph. Model.} \textbf{\bibinfo{volume}{21}},
  \bibinfo{pages}{289} (\bibinfo{year}{2003}).

\bibitem[{\citenamefont{Diller and Verlinde}(1999)}]{Diller1999}
\bibinfo{author}{\bibfnamefont{D.~J.} \bibnamefont{Diller}} \bibnamefont{and}
  \bibinfo{author}{\bibfnamefont{C.~L. M.~J.} \bibnamefont{Verlinde}},
  \bibinfo{journal}{J. Comput. Chem.} \textbf{\bibinfo{volume}{20}},
  \bibinfo{pages}{1740} (\bibinfo{year}{1999}).

\bibitem[{\citenamefont{Michel and Essex}(2010)}]{Michel2010}
\bibinfo{author}{\bibfnamefont{J.}~\bibnamefont{Michel}} \bibnamefont{and}
  \bibinfo{author}{\bibfnamefont{J.~W.} \bibnamefont{Essex}},
  \bibinfo{journal}{J. Comput.-Aided Mol. Des.} \textbf{\bibinfo{volume}{24}},
  \bibinfo{pages}{639} (\bibinfo{year}{2010}).

\bibitem[{\citenamefont{Duane et~al.}(1987)\citenamefont{Duane, Kennedy,
  Pendleton, and Roweth}}]{Duane1987}
\bibinfo{author}{\bibfnamefont{S.}~\bibnamefont{Duane}},
  \bibinfo{author}{\bibfnamefont{A.~D.} \bibnamefont{Kennedy}},
  \bibinfo{author}{\bibfnamefont{B.~J.} \bibnamefont{Pendleton}},
  \bibnamefont{and} \bibinfo{author}{\bibfnamefont{D.}~\bibnamefont{Roweth}},
  \bibinfo{journal}{Phys. Lett. B} \textbf{\bibinfo{volume}{195}},
  \bibinfo{pages}{216} (\bibinfo{year}{1987}).

\bibitem[{\citenamefont{Jiang and Roux}(2010)}]{Jiang2010}
\bibinfo{author}{\bibfnamefont{W.}~\bibnamefont{Jiang}} \bibnamefont{and}
  \bibinfo{author}{\bibfnamefont{B.}~\bibnamefont{Roux}}, \bibinfo{journal}{J.
  Chem. Theory Comput.} \textbf{\bibinfo{volume}{6}}, \bibinfo{pages}{2559}
  (\bibinfo{year}{2010}).

\bibitem[{\citenamefont{Chodera and Shirts}(2011)}]{Chodera2011a}
\bibinfo{author}{\bibfnamefont{J.~D.} \bibnamefont{Chodera}} \bibnamefont{and}
  \bibinfo{author}{\bibfnamefont{M.~R.} \bibnamefont{Shirts}},
  \bibinfo{journal}{J. Chem. Phys.} \textbf{\bibinfo{volume}{135}},
  \bibinfo{pages}{194110} (\bibinfo{year}{2011}).

\bibitem[{\citenamefont{Lu and Kofke}(2001)}]{Lu2001}
\bibinfo{author}{\bibfnamefont{N.}~\bibnamefont{Lu}} \bibnamefont{and}
  \bibinfo{author}{\bibfnamefont{D.~A.} \bibnamefont{Kofke}},
  \bibinfo{journal}{J. Chem. Phys.} \textbf{\bibinfo{volume}{114}},
  \bibinfo{pages}{7303} (\bibinfo{year}{2001}).

\bibitem[{\citenamefont{Weinhold}(1975)}]{Weinhold1975}
\bibinfo{author}{\bibfnamefont{F.}~\bibnamefont{Weinhold}},
  \bibinfo{journal}{J. Chem. Phys.} \textbf{\bibinfo{volume}{63}},
  \bibinfo{pages}{2479} (\bibinfo{year}{1975}).

\bibitem[{\citenamefont{Shenfeld et~al.}(2009)\citenamefont{Shenfeld, Xu,
  Eastwood, Dror, and Shaw}}]{Shenfeld2009}
\bibinfo{author}{\bibfnamefont{D.~K.} \bibnamefont{Shenfeld}},
  \bibinfo{author}{\bibfnamefont{H.}~\bibnamefont{Xu}},
  \bibinfo{author}{\bibfnamefont{M.~P.} \bibnamefont{Eastwood}},
  \bibinfo{author}{\bibfnamefont{R.~O.} \bibnamefont{Dror}}, \bibnamefont{and}
  \bibinfo{author}{\bibfnamefont{D.~E.} \bibnamefont{Shaw}},
  \bibinfo{journal}{Phys. Rev. E} \textbf{\bibinfo{volume}{80}},
  \bibinfo{pages}{46705} (\bibinfo{year}{2009}).

\bibitem[{\citenamefont{Crooks}(2007)}]{Crooks2007}
\bibinfo{author}{\bibfnamefont{G.~E.} \bibnamefont{Crooks}},
  \bibinfo{journal}{Phys. Rev. Lett.} \textbf{\bibinfo{volume}{99}},
  \bibinfo{pages}{100602} (\bibinfo{year}{2007}).

\bibitem[{\citenamefont{Nguyen and Minh}(2016)}]{Nguyen2016}
\bibinfo{author}{\bibfnamefont{T.~H.} \bibnamefont{Nguyen}} \bibnamefont{and}
  \bibinfo{author}{\bibfnamefont{D.~D.~L.} \bibnamefont{Minh}},
  \bibinfo{journal}{J. Chem. Theory Comput.} \textbf{\bibinfo{volume}{12}},
  \bibinfo{pages}{2154} (\bibinfo{year}{2016}).

\bibitem[{\citenamefont{Chodera}(2016)}]{Chodera2016}
\bibinfo{author}{\bibfnamefont{J.~D.} \bibnamefont{Chodera}},
  \bibinfo{journal}{J. Chem. Theory Comput.} \textbf{\bibinfo{volume}{12}},
  \bibinfo{pages}{1799} (\bibinfo{year}{2016}).

\bibitem[{\citenamefont{Zwanzig}(1954)}]{Zwanzig1954}
\bibinfo{author}{\bibfnamefont{R.}~\bibnamefont{Zwanzig}}, \bibinfo{journal}{J.
  Chem. Phys.} \textbf{\bibinfo{volume}{22}}, \bibinfo{pages}{1420}
  (\bibinfo{year}{1954}).

\bibitem[{\citenamefont{{van der Walt} et~al.}(2011)\citenamefont{{van der
  Walt}, Colbert, and Varoquaux}}]{Walt2011numpy}
\bibinfo{author}{\bibfnamefont{S.}~\bibnamefont{{van der Walt}}},
  \bibinfo{author}{\bibfnamefont{S.~C.} \bibnamefont{Colbert}},
  \bibnamefont{and}
  \bibinfo{author}{\bibfnamefont{G.}~\bibnamefont{Varoquaux}},
  \bibinfo{journal}{Comput. Sci. Eng.} \textbf{\bibinfo{volume}{13}},
  \bibinfo{pages}{22} (\bibinfo{year}{2011}).

\bibitem[{\citenamefont{Lang et~al.}(2009)\citenamefont{Lang, Brozell,
  Mukherjee, Pettersen, Meng, Thomas, Rizzo, Case, James, and
  Kuntz}}]{Lang2009}
\bibinfo{author}{\bibfnamefont{P.}~\bibnamefont{Lang}},
  \bibinfo{author}{\bibfnamefont{S.~R.} \bibnamefont{Brozell}},
  \bibinfo{author}{\bibfnamefont{S.}~\bibnamefont{Mukherjee}},
  \bibinfo{author}{\bibfnamefont{E.}~\bibnamefont{Pettersen}},
  \bibinfo{author}{\bibfnamefont{E.~C.} \bibnamefont{Meng}},
  \bibinfo{author}{\bibfnamefont{V.}~\bibnamefont{Thomas}},
  \bibinfo{author}{\bibfnamefont{R.~C.} \bibnamefont{Rizzo}},
  \bibinfo{author}{\bibfnamefont{D.~A.} \bibnamefont{Case}},
  \bibinfo{author}{\bibfnamefont{T.}~\bibnamefont{James}}, \bibnamefont{and}
  \bibinfo{author}{\bibfnamefont{I.~D.} \bibnamefont{Kuntz}},
  \bibinfo{journal}{RNA} \textbf{\bibinfo{volume}{15}}, \bibinfo{pages}{1219}
  (\bibinfo{year}{2009}).

\bibitem[{\citenamefont{Pordes et~al.}(2007)\citenamefont{Pordes, Petravick,
  Kramer, Olson, Livny, Roy, Avery, Blackburn, Wenaus, W{\"u}rthwein
  et~al.}}]{Pordes2007}
\bibinfo{author}{\bibfnamefont{R.}~\bibnamefont{Pordes}},
  \bibinfo{author}{\bibfnamefont{D.}~\bibnamefont{Petravick}},
  \bibinfo{author}{\bibfnamefont{B.}~\bibnamefont{Kramer}},
  \bibinfo{author}{\bibfnamefont{D.}~\bibnamefont{Olson}},
  \bibinfo{author}{\bibfnamefont{M.}~\bibnamefont{Livny}},
  \bibinfo{author}{\bibfnamefont{A.}~\bibnamefont{Roy}},
  \bibinfo{author}{\bibfnamefont{P.}~\bibnamefont{Avery}},
  \bibinfo{author}{\bibfnamefont{K.}~\bibnamefont{Blackburn}},
  \bibinfo{author}{\bibfnamefont{T.}~\bibnamefont{Wenaus}},
  \bibinfo{author}{\bibfnamefont{F.}~\bibnamefont{W{\"u}rthwein}},
  \bibnamefont{et~al.}, \bibinfo{journal}{J. Phys.: Conf. Ser.}
  \textbf{\bibinfo{volume}{78}}, \bibinfo{pages}{012057}
  (\bibinfo{year}{2007}).

\bibitem[{\citenamefont{Towns et~al.}(Sept.-Oct. 2014)\citenamefont{Towns,
  Cockerill, Dahan, Foster, Gaither, Grimshaw, Hazlewood, Lathrop, Lifka,
  Peterson et~al.}}]{xsede}
\bibinfo{author}{\bibfnamefont{J.}~\bibnamefont{Towns}},
  \bibinfo{author}{\bibfnamefont{T.}~\bibnamefont{Cockerill}},
  \bibinfo{author}{\bibfnamefont{M.}~\bibnamefont{Dahan}},
  \bibinfo{author}{\bibfnamefont{I.}~\bibnamefont{Foster}},
  \bibinfo{author}{\bibfnamefont{K.}~\bibnamefont{Gaither}},
  \bibinfo{author}{\bibfnamefont{A.}~\bibnamefont{Grimshaw}},
  \bibinfo{author}{\bibfnamefont{V.}~\bibnamefont{Hazlewood}},
  \bibinfo{author}{\bibfnamefont{S.}~\bibnamefont{Lathrop}},
  \bibinfo{author}{\bibfnamefont{D.}~\bibnamefont{Lifka}},
  \bibinfo{author}{\bibfnamefont{G.~D.} \bibnamefont{Peterson}},
  \bibnamefont{et~al.}, \bibinfo{journal}{Comput. Sci. Eng.}
  \textbf{\bibinfo{volume}{16}}, \bibinfo{pages}{62} (\bibinfo{year}{Sept.-Oct.
  2014}).

\bibitem[{\citenamefont{Wood et~al.}(1991)\citenamefont{Wood, Muhlbauer,
  Thompson, M{\"u}hlbauer, and Thompson}}]{Wood1991}
\bibinfo{author}{\bibfnamefont{R.~H.} \bibnamefont{Wood}},
  \bibinfo{author}{\bibfnamefont{W.~C.~F.} \bibnamefont{Muhlbauer}},
  \bibinfo{author}{\bibfnamefont{P.~T.} \bibnamefont{Thompson}},
  \bibinfo{author}{\bibfnamefont{W.~C.} \bibnamefont{M{\"u}hlbauer}},
  \bibnamefont{and} \bibinfo{author}{\bibfnamefont{P.~T.}
  \bibnamefont{Thompson}}, \bibinfo{journal}{J. Phys. Chem.}
  \textbf{\bibinfo{volume}{95}}, \bibinfo{pages}{6670} (\bibinfo{year}{1991}).

\bibitem[{\citenamefont{Mobley}(2012)}]{Mobley2012b}
\bibinfo{author}{\bibfnamefont{D.~L.} \bibnamefont{Mobley}},
  \bibinfo{journal}{J. Comput.-Aided Mol. Des.} \textbf{\bibinfo{volume}{26}},
  \bibinfo{pages}{93} (\bibinfo{year}{2012}).

\bibitem[{\citenamefont{Mysinger and Shoichet}(2010)}]{Mysinger2010}
\bibinfo{author}{\bibfnamefont{M.~M.} \bibnamefont{Mysinger}} \bibnamefont{and}
  \bibinfo{author}{\bibfnamefont{B.~K.} \bibnamefont{Shoichet}},
  \bibinfo{journal}{J. Chem. Inf. Model.} \textbf{\bibinfo{volume}{50}},
  \bibinfo{pages}{1561} (\bibinfo{year}{2010}).

\bibitem[{\citenamefont{Repasky et~al.}(2012)\citenamefont{Repasky, Murphy,
  Banks, Greenwood, {Tubert-Brohman}, Bhat, and Friesner}}]{Repasky2012}
\bibinfo{author}{\bibfnamefont{M.~P.} \bibnamefont{Repasky}},
  \bibinfo{author}{\bibfnamefont{R.~B.} \bibnamefont{Murphy}},
  \bibinfo{author}{\bibfnamefont{J.~L.} \bibnamefont{Banks}},
  \bibinfo{author}{\bibfnamefont{J.~R.} \bibnamefont{Greenwood}},
  \bibinfo{author}{\bibfnamefont{I.}~\bibnamefont{{Tubert-Brohman}}},
  \bibinfo{author}{\bibfnamefont{S.}~\bibnamefont{Bhat}}, \bibnamefont{and}
  \bibinfo{author}{\bibfnamefont{R.~A.} \bibnamefont{Friesner}},
  \bibinfo{journal}{J. Comput.-Aided Mol. Des.} \textbf{\bibinfo{volume}{26}},
  \bibinfo{pages}{787} (\bibinfo{year}{2012}).

\bibitem[{\citenamefont{Neves et~al.}(2012)\citenamefont{Neves, Totrov, and
  Abagyan}}]{Neves2012}
\bibinfo{author}{\bibfnamefont{M.~a.~C.} \bibnamefont{Neves}},
  \bibinfo{author}{\bibfnamefont{M.}~\bibnamefont{Totrov}}, \bibnamefont{and}
  \bibinfo{author}{\bibfnamefont{R.~A.} \bibnamefont{Abagyan}},
  \bibinfo{journal}{J. Comput.-Aided Mol. Des.} \textbf{\bibinfo{volume}{26}},
  \bibinfo{pages}{675} (\bibinfo{year}{2012}).

\bibitem[{\citenamefont{Wang et~al.}(2016)\citenamefont{Wang, Sun, Yao, Li, Xu,
  Li, Tian, and Hou}}]{Wang2016d}
\bibinfo{author}{\bibfnamefont{Z.}~\bibnamefont{Wang}},
  \bibinfo{author}{\bibfnamefont{H.}~\bibnamefont{Sun}},
  \bibinfo{author}{\bibfnamefont{X.}~\bibnamefont{Yao}},
  \bibinfo{author}{\bibfnamefont{D.}~\bibnamefont{Li}},
  \bibinfo{author}{\bibfnamefont{L.}~\bibnamefont{Xu}},
  \bibinfo{author}{\bibfnamefont{Y.}~\bibnamefont{Li}},
  \bibinfo{author}{\bibfnamefont{S.}~\bibnamefont{Tian}}, \bibnamefont{and}
  \bibinfo{author}{\bibfnamefont{T.}~\bibnamefont{Hou}},
  \bibinfo{journal}{Phys. Chem. Chem. Phys.} \textbf{\bibinfo{volume}{18}},
  \bibinfo{pages}{12964} (\bibinfo{year}{2016}).

\bibitem[{\citenamefont{Sfiligoi et~al.}(2009)\citenamefont{Sfiligoi, Bradley,
  Holzman, Mhashilkar, Padhi, and Wurthwein}}]{Sfiligoi2009}
\bibinfo{author}{\bibfnamefont{I.}~\bibnamefont{Sfiligoi}},
  \bibinfo{author}{\bibfnamefont{D.~C.} \bibnamefont{Bradley}},
  \bibinfo{author}{\bibfnamefont{B.}~\bibnamefont{Holzman}},
  \bibinfo{author}{\bibfnamefont{P.}~\bibnamefont{Mhashilkar}},
  \bibinfo{author}{\bibfnamefont{S.}~\bibnamefont{Padhi}}, \bibnamefont{and}
  \bibinfo{author}{\bibfnamefont{F.}~\bibnamefont{Wurthwein}}, in
  \emph{\bibinfo{booktitle}{2009 WRI World Congress on Computer Science and
  Information Engineering}} (\bibinfo{publisher}{IEEE}, \bibinfo{address}{Los
  Angeles, California USA}, \bibinfo{year}{2009}), pp.
  \bibinfo{pages}{428--432}, ISBN \bibinfo{isbn}{978-0-7695-3507-4}.

\bibitem[{\citenamefont{Humphrey et~al.}(1996)\citenamefont{Humphrey, Dalke,
  and Schulten}}]{Humphrey1996}
\bibinfo{author}{\bibfnamefont{W.}~\bibnamefont{Humphrey}},
  \bibinfo{author}{\bibfnamefont{A.}~\bibnamefont{Dalke}}, \bibnamefont{and}
  \bibinfo{author}{\bibfnamefont{K.}~\bibnamefont{Schulten}},
  \bibinfo{journal}{J. Mol. Graphics} \textbf{\bibinfo{volume}{14}},
  \bibinfo{pages}{33} (\bibinfo{year}{1996}).

\end{thebibliography}

\clearpage

\setcounter{figure}{0}
\makeatletter
\renewcommand{\thefigure}{S\arabic{figure}}
\renewcommand{\thetable}{S\arabic{table}}
\setcounter{equation}{0}
\renewcommand{\theequation}{S\arabic{equation}}

\begin{figure}[p]
\begin{center}
\includegraphics[width=6.65in]{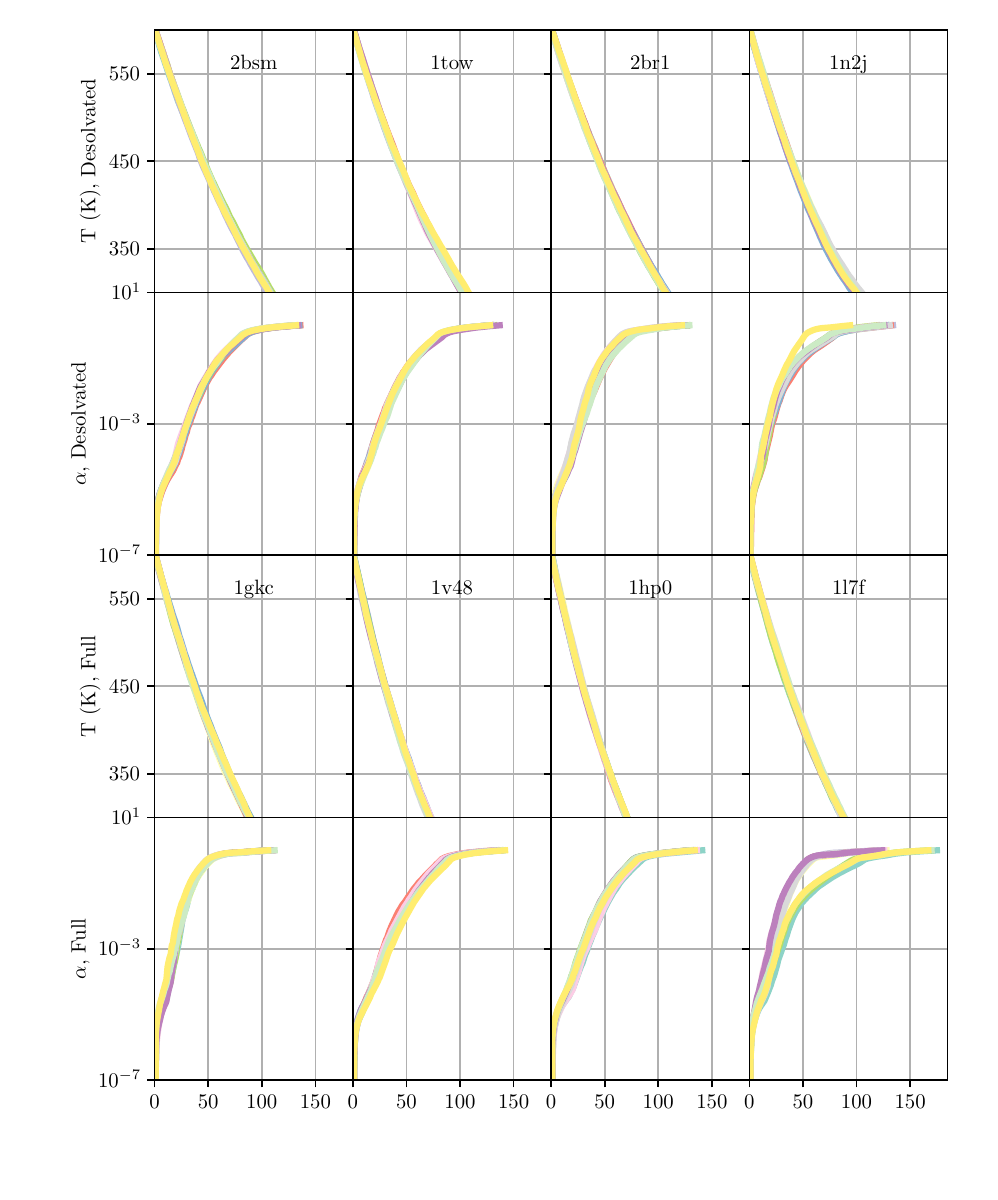}
\end{center}
\caption{
{\bf Protocols} 
for 11 independent simulations.
In the top two rows, the columns contain protocols for the Desolvated option 
for 2bsm, 1tow, 2br1, and 1n2j (from left to right).
For these systems, $\sigma[N_{states}]$ for states CD is 1.85, 2.56, 3.30, and 10.5, respectively.
In the bottom two rows, the columns contain protocols for the Full option 
for 1gkc, 1v48, 1hp0, and 1l7f, respectively.
For these systems, $\sigma[N_{states}]$ for states CD is 1.91, 2.42, 3.34, and 22.4, respectively.
\label{fig:protocols}}
\end{figure}

\clearpage

\begin{figure}[p]
\begin{center}
\includegraphics[width=3.18in]{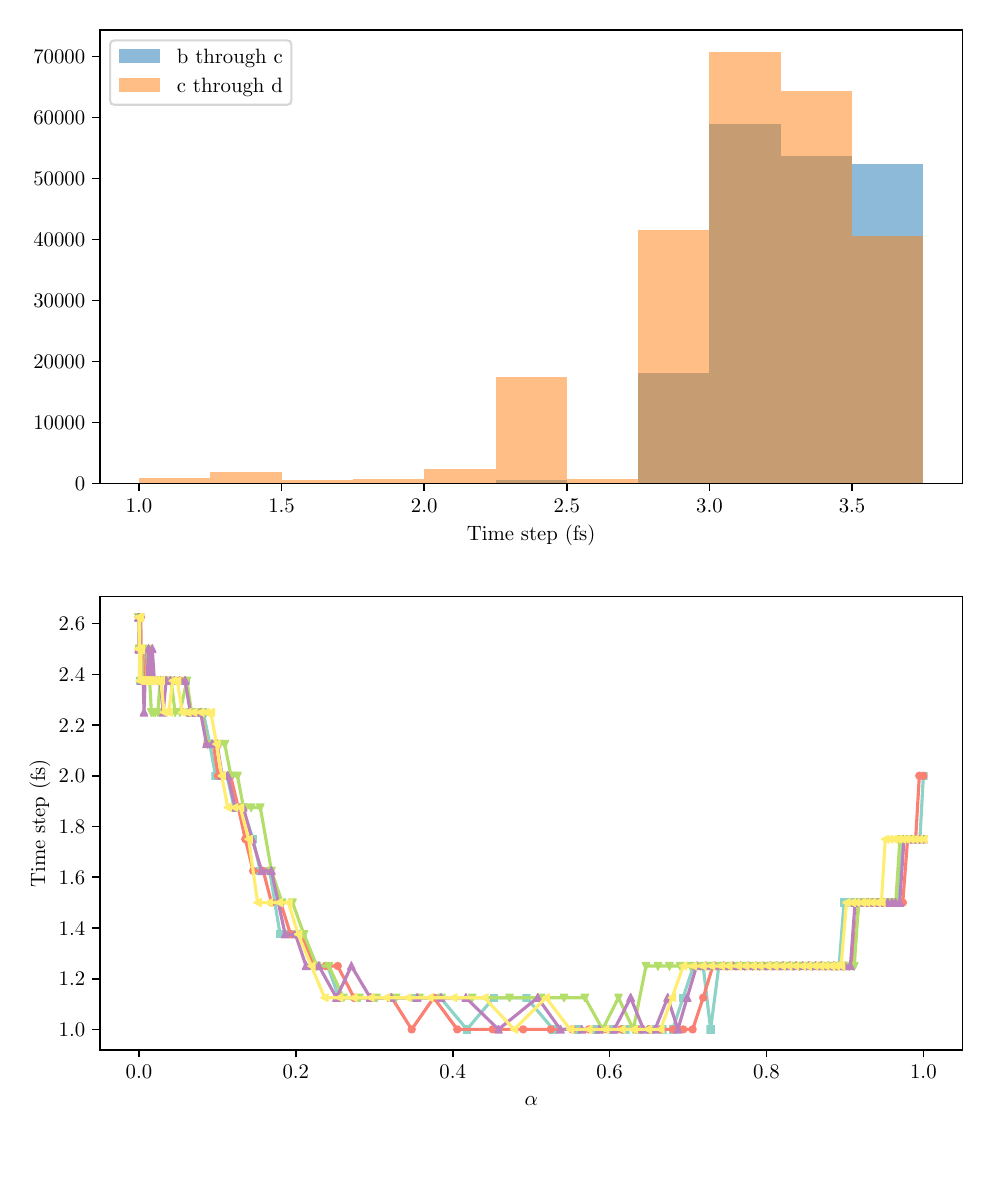}
\end{center}
\caption{
{\bf Time step statistics.} 
Top: Histograms of time steps for states BC and states CD.
Bottom: Five protocols with the shortest observed time steps 
for states CD.
These protocols are from simulations with 1lrh.
\label{fig:time_steps}}
\end{figure}

\clearpage

\begin{figure}[p]
\begin{center}
\includegraphics[width=6.64in]{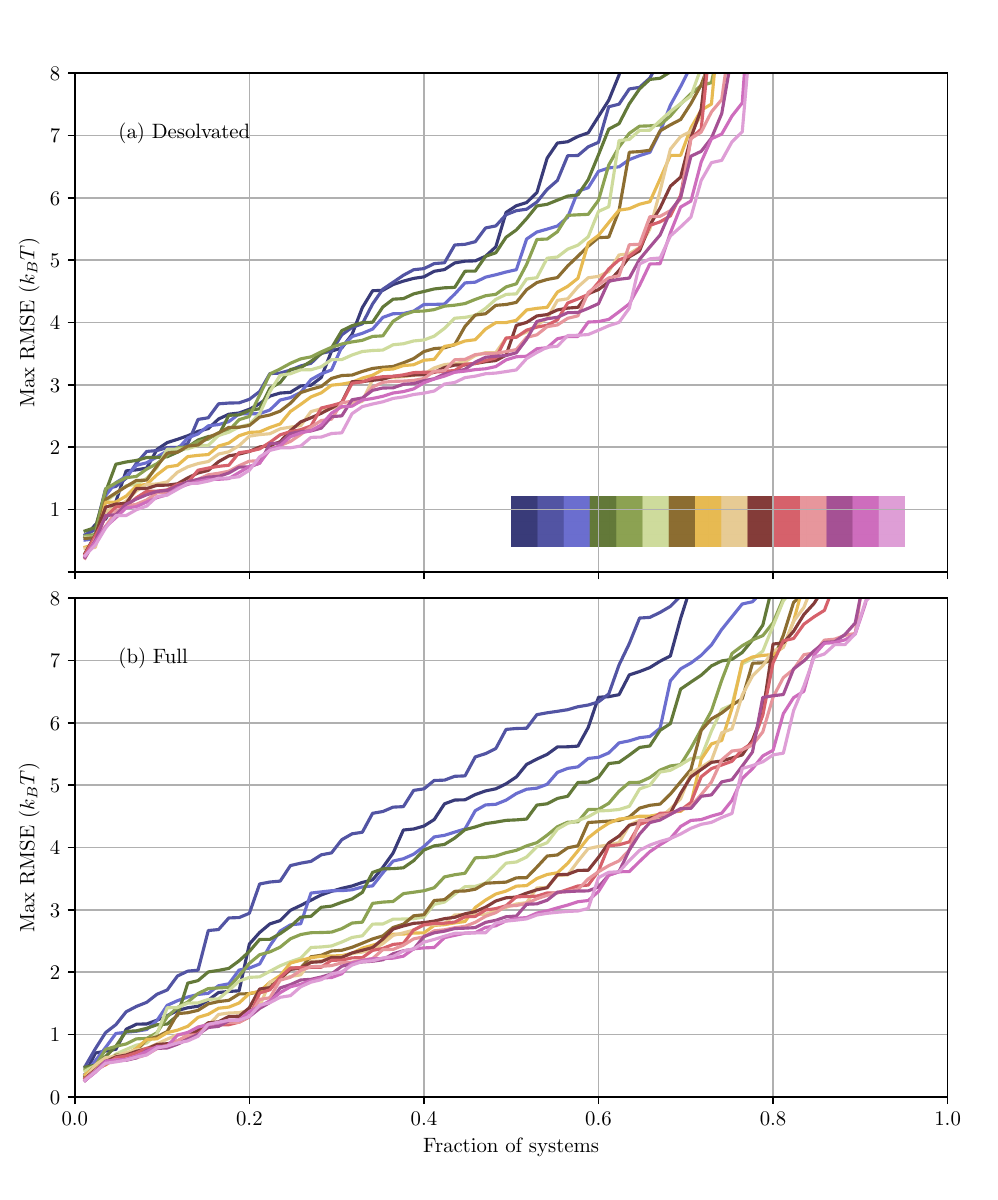}
\end{center}
\caption{
{\bf Fraction of systems} with free energy differences calculated within a certain root mean square error of the consensus value - the lower of the mean BPMFs for each solvation option . Each line indicates a different number of cycles, with a total 15. In the sequence of colors inset in the top panel, the color indicates an increasing number of cycles from left to right.
\label{fig:corrected_convergence}}
\end{figure}

\clearpage

\begin{figure}[p]
\begin{center}
\includegraphics[width=3.18in]{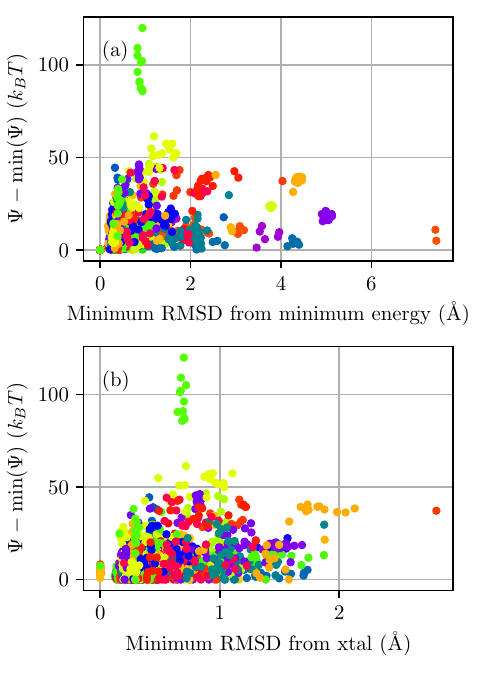}
\end{center}
\caption{
{\bf Differences in the interaction energy} as a function of the minimum observed RMSD (a) from the structure with the lowest interaction energy or (b) from the crystal structure. The interaction energy is based on the force field in milestone E.
\label{fig:distance_vs_interaction}}
\end{figure}

\clearpage

\begin{center}
\begin{longtable}{| l | c c c c | c |}
\hline \hline
\multicolumn{6}{c}{Desolvated} \\
\hline \hline
PDB ID & $f_{AB}$ & $f_{BC}$ & $f_{CD}$ & $f_{DE}$ & $f_{AE}$ \\ 
\hline 
1jla &   62.88 (0.04) & -154.13 (0.06) & -137.10 (0.15) &   -3.46 (0.21) &  -49.31 (0.26) \\ 
1tz8 &   23.80 (0.02) &    3.33 (0.05) &   39.84 (0.17) &   -7.12 (0.32) &    5.58 (0.38) \\ 
1sqn &   28.46 (0.06) &   20.18 (0.05) &    2.79 (0.08) &  -17.37 (0.38) &  -63.21 (0.43) \\ 
1w2g &   76.64 (0.23) & -188.70 (0.04) & -136.14 (0.32) &  -31.68 (0.78) &  -55.77 (0.61) \\ 
1n46 &   42.40 (0.04) &  -88.19 (0.06) &   -7.34 (0.19) & -118.08 (0.61) &  -79.64 (0.70) \\ 
1r9o &   30.14 (0.02) & -117.83 (0.03) &    5.28 (0.29) & -129.18 (0.71) &  -36.21 (0.77) \\ 
1opk &   96.89 (0.07) & -287.25 (0.04) & -247.78 (0.46) &  -26.04 (0.62) &  -83.46 (0.78) \\ 
1nav &   38.05 (0.01) & -136.52 (0.06) &  -44.69 (0.13) &  -95.42 (0.86) &  -41.63 (0.89) \\ 
1y6b &  118.25 (0.07) & -226.39 (0.08) &  -72.01 (0.53) &  -94.72 (0.99) &  -58.59 (0.90) \\ 
1q41 &   58.36 (0.06) &  -30.06 (0.05) &   11.10 (0.36) &  -19.94 (0.93) &  -37.14 (0.96) \\ 
1ke5 &  127.48 (0.06) & -231.33 (0.04) & -116.11 (0.47) &  -17.48 (1.04) &  -29.73 (1.02) \\ 
1xoz &   64.97 (0.11) &   13.27 (0.03) &   38.08 (0.62) &  -19.59 (0.69) &  -59.75 (1.09) \\ 
2br1 &   90.57 (0.07) & -207.78 (0.05) & -144.88 (0.75) &   -0.44 (1.00) &  -28.10 (1.15) \\ 
1q4g &   29.51 (0.02) & -116.52 (0.04) &  -12.60 (0.13) & -124.02 (1.15) &  -49.61 (1.19) \\ 
1gpk &   36.57 (0.04) & -142.54 (0.05) &  -90.21 (0.12) &  -54.08 (1.18) &  -38.32 (1.21) \\ 
1w1p &   46.64 (0.17) &   -4.50 (0.04) &   41.74 (0.09) &  -10.99 (1.22) &  -11.39 (1.23) \\ 
1z95 &  110.55 (0.04) & -152.57 (0.14) &  -60.24 (0.21) &  -31.50 (1.05) &  -49.73 (1.23) \\ 
1pmn &   80.88 (0.05) & -193.64 (0.35) &  -88.21 (1.36) &  -88.33 (1.27) &  -63.78 (1.42) \\ 
1sg0 &   33.32 (0.02) &  -86.30 (0.04) &  -77.49 (0.69) &  -10.83 (1.25) &  -35.35 (1.43) \\ 
1yv3 &   52.46 (0.08) &  -48.62 (0.06) &  -32.31 (0.08) &  -20.24 (1.46) &  -56.38 (1.48) \\ 
1kzk &   97.80 (0.16) &  -55.14 (0.12) &   -4.10 (0.43) &   -1.48 (1.33) &  -48.24 (1.51) \\ 
1owe &   34.50 (0.11) & -114.98 (0.06) &  -32.13 (0.28) &  -89.24 (1.52) &  -40.89 (1.52) \\ 
2bm2 &   52.93 (0.05) & -101.49 (0.20) &  -30.88 (0.95) &  -92.13 (1.39) &  -74.45 (1.63) \\ 
1of1 &   70.31 (0.09) & -162.72 (0.06) & -109.79 (0.53) &  -38.61 (1.80) &  -55.99 (1.68) \\ 
1n2v &   56.68 (0.05) &   -1.48 (0.04) &   45.16 (0.31) &  -17.67 (1.84) &  -27.71 (1.75) \\ 
1hq2 &  103.15 (0.10) & -294.03 (0.02) & -195.67 (0.09) &  -21.26 (1.75) &  -26.06 (1.79) \\ 
1u1c &   68.85 (0.06) & -173.23 (0.04) & -126.22 (0.54) &  -32.22 (1.96) &  -54.06 (1.84) \\ 
1v4s &   86.02 (0.09) & -196.64 (0.03) & -142.55 (0.47) &  -10.46 (1.92) &  -42.39 (1.90) \\ 
1yvf &   54.96 (0.04) & -105.44 (0.04) &   16.50 (0.43) &  -95.38 (1.93) &  -28.40 (1.93) \\ 
1hvy &   86.46 (0.11) & -446.19 (0.05) &  -84.75 (0.72) & -334.35 (1.90) &  -59.36 (1.95) \\ 
1ia1 &   84.01 (0.10) & -332.47 (0.04) & -209.62 (0.78) &  -55.36 (1.80) &  -16.53 (1.98) \\ 
1t46 &  117.93 (0.09) & -291.58 (0.10) & -146.42 (0.54) & -133.97 (1.71) & -106.73 (2.02) \\ 
1xoq &   54.16 (0.05) &  -39.40 (0.05) &  -23.24 (1.20) &   23.84 (1.63) &  -14.17 (2.04) \\ 
1ygc &  150.59 (0.14) & -375.77 (0.06) & -200.64 (1.06) &  -74.85 (2.08) &  -50.30 (2.08) \\ 
1gkc &   58.36 (0.13) & -112.91 (0.13) &   26.39 (0.96) & -103.06 (2.21) &  -22.12 (2.12) \\ 
1gm8 &   78.20 (0.16) & -134.63 (0.04) &   20.85 (1.61) &  -99.79 (1.52) &  -22.51 (2.20) \\ 
1v48 &   73.86 (0.12) & -436.43 (0.18) &  -20.11 (1.15) & -419.11 (2.20) &  -76.64 (2.22) \\ 
1m2z &   59.35 (0.07) &   59.12 (0.05) &   70.99 (0.12) &  -21.88 (2.23) &  -69.36 (2.23) \\ 
1k3u &   50.01 (0.08) & -330.64 (0.14) &   61.13 (0.29) & -387.62 (2.35) &  -45.86 (2.28) \\ 
2bsm &   55.93 (0.04) &  -41.58 (0.04) &   -1.53 (1.61) &  -19.87 (1.75) &  -35.77 (2.32) \\ 
1j3j &   80.18 (0.11) & -325.93 (0.04) & -217.48 (0.46) &  -55.37 (2.20) &  -27.09 (2.37) \\ 
1sj0 &   29.00 (0.04) &  -24.47 (0.07) &    7.83 (0.40) &  -84.58 (2.35) &  -81.29 (2.43) \\ 
1tow &   36.50 (0.03) & -124.24 (0.04) &  -17.44 (0.27) &  -84.63 (2.34) &  -14.34 (2.44) \\ 
1meh &   61.51 (0.06) & -166.72 (0.09) &  -21.27 (0.94) & -109.34 (2.20) &  -25.39 (2.54) \\ 
1s19 &   39.48 (0.03) &   38.94 (0.13) &   21.90 (0.21) &  -24.29 (2.54) &  -80.80 (2.59) \\ 
1unl &  133.70 (0.09) & -372.20 (0.07) & -276.15 (1.85) &    5.90 (1.66) &  -31.75 (2.65) \\ 
1v0p &  145.17 (0.08) & -473.13 (0.07) & -264.08 (1.07) & -110.90 (2.03) &  -47.02 (2.73) \\ 
1n1m &   28.73 (0.08) &  -35.44 (0.04) &   56.22 (0.34) & -124.52 (2.85) &  -61.59 (2.87) \\ 
1ywr &  102.60 (0.07) & -252.81 (0.16) & -114.01 (0.84) & -104.88 (2.84) &  -68.68 (2.90) \\ 
1l2s &  108.62 (0.04) & -268.55 (0.05) &  -95.93 (0.31) & -123.52 (2.98) &  -59.51 (3.01) \\ 
1g9v &   53.63 (0.07) & -125.07 (0.04) &    1.70 (0.22) &  -86.80 (3.07) &  -13.66 (3.11) \\ 
1oyt &   63.79 (0.18) & -152.57 (0.07) &  -63.60 (2.33) &  -73.04 (2.41) &  -47.87 (3.12) \\ 
1u4d &  104.71 (0.10) & -229.11 (0.05) & -119.44 (0.34) &  -33.69 (3.21) &  -28.74 (3.24) \\ 
1lpz &   45.77 (0.13) &  -96.57 (0.15) &  -23.63 (0.65) &  -70.17 (3.55) &  -43.00 (3.25) \\ 
1of6 &   72.62 (0.02) & -109.25 (0.03) &    1.66 (0.09) &  -99.76 (3.30) &  -61.47 (3.27) \\ 
1hnn &   91.00 (0.02) & -243.03 (0.03) & -114.06 (0.16) &  -49.46 (3.51) &  -11.48 (3.45) \\ 
1hww &   26.16 (0.02) &    1.34 (0.03) &   91.52 (0.31) &  -68.20 (3.37) &   -4.19 (3.54) \\ 
1t9b &  224.57 (0.04) & -866.45 (0.16) & -598.58 (1.46) &  -81.82 (3.16) &  -38.52 (3.54) \\ 
1ig3 &   70.94 (0.08) & -307.74 (0.03) & -206.32 (0.86) &  -55.48 (3.38) &  -24.99 (3.59) \\ 
1s3v &  100.80 (0.12) & -347.86 (0.29) & -223.05 (1.42) &  -86.81 (2.68) &  -62.80 (3.60) \\ 
1uml &   87.78 (0.09) & -104.06 (0.15) &  -31.75 (0.93) &   15.37 (3.49) &   -0.10 (3.88) \\ 
1sq5 &   67.62 (0.06) & -174.34 (0.18) &  -16.20 (0.11) &  -81.53 (3.78) &    8.98 (3.89) \\ 
1p62 &  101.64 (0.54) & -226.64 (0.04) & -122.73 (0.11) &  -14.16 (3.98) &  -11.89 (3.94) \\ 
1lrh &   28.56 (0.02) & -131.13 (0.05) &  -69.61 (0.17) & -115.55 (4.00) &  -82.59 (3.95) \\ 
1tt1 &   83.00 (0.02) & -180.49 (0.11) &   26.39 (0.09) & -127.45 (4.10) &   -3.58 (4.08) \\ 
1vcj &  111.49 (0.12) & -171.89 (0.05) &   -9.44 (0.85) & -105.75 (4.92) &  -54.79 (5.04) \\ 
1yqy &   91.67 (0.04) & -191.96 (0.11) &  -12.54 (0.41) & -147.67 (5.03) &  -59.91 (5.13) \\ 
1mmv &   67.30 (0.04) & -139.64 (0.14) &  -36.52 (2.28) &  -22.33 (7.26) &   13.50 (5.79) \\ 
1p2y &   16.47 (0.01) &  -38.46 (0.05) &  -45.16 (0.08) &   -8.99 (5.89) &  -32.16 (5.89) \\ 
1jje &   26.49 (0.04) & -288.84 (0.99) &  -94.34 (2.16) & -252.19 (7.07) &  -84.18 (6.60) \\ 
1l7f &  106.47 (0.11) & -264.01 (0.07) & -127.69 (1.03) &  -81.92 (6.03) &  -52.07 (6.63) \\ 
1mzc &   60.01 (0.11) & -104.79 (0.13) &  -16.04 (0.78) &  -89.92 (6.68) &  -61.17 (6.83) \\ 
1q1g &   64.45 (0.05) & -150.25 (0.05) &  -72.51 (0.38) &  -59.55 (7.43) &  -46.26 (7.38) \\ 
1oq5 &  111.71 (0.03) & -175.90 (0.05) &  -12.35 (0.37) & -175.52 (7.95) & -123.68 (7.90) \\ 
1n2j &   44.11 (0.03) &  -98.37 (0.09) &   30.03 (0.24) &  -92.94 (7.94) &   -8.65 (8.08) \\ 
1t40 &   64.80 (0.05) & -148.84 (0.04) &  -58.19 (2.65) &  -81.11 (7.07) &  -55.26 (8.09) \\ 
1x8x &   72.62 (0.02) & -109.25 (0.04) &    1.37 (0.28) &  -66.80 (8.65) &  -28.80 (8.56) \\ 
1hp0 &  106.81 (0.31) & -162.12 (0.04) &  -69.65 (0.35) &   17.10 (8.93) &    2.76 (8.83) \\ 
1jd0 &  117.56 (0.05) & -256.48 (0.07) &  -46.98 (0.67) & -142.24 (9.08) &  -50.30 (9.07) \\ 
1r58 &   73.43 (0.05) &  -19.93 (0.08) &   32.82 (0.66) &   44.37 (9.93) &   23.69 (9.84) \\ 
1r1h &  132.40 (0.07) & -216.80 (0.65) &    6.60 (1.17) & -147.44 (10.92) &  -56.43 (10.87) \\ 
1uou &   67.51 (0.09) & -227.43 (0.04) & -138.42 (0.17) &  -46.25 (11.22) &  -24.75 (11.33) \\ 
1r55 &   64.41 (0.11) &  -95.71 (0.31) &   41.62 (0.89) &  -86.67 (11.90) &  -13.76 (12.04) \\ 
1xm6 &   39.35 (0.07) &  -26.42 (0.04) &  -43.50 (2.13) &  207.94 (11.52) &  151.51 (12.64) \\ 
1hwi &   61.71 (0.03) & -170.89 (0.10) &  -43.77 (0.89) &  -85.85 (19.61) &  -20.44 (19.21) \\ 
\hline \hline
\multicolumn{6}{c}{Full} \\
\hline \hline
PDB ID & $f_{AB}$ & $f_{BC}$ & $f_{CD}$ & $f_{DE}$ & $f_{AE}$ \\ 
\hline 
1sqn &   28.47 (0.04) &   40.65 (0.07) &  -17.42 (0.11) &   23.67 (0.25) &  -62.88 (0.22) \\ 
1jla &   62.90 (0.08) & -111.67 (0.05) & -177.93 (0.17) &   79.81 (0.14) &  -49.36 (0.27) \\ 
1yv3 &   52.44 (0.10) &  -11.54 (0.06) &  -68.20 (0.11) &   51.39 (0.51) &  -57.71 (0.54) \\ 
1yvf &   54.97 (0.06) &  -10.99 (0.05) &  -72.32 (0.42) &   86.95 (0.27) &  -29.36 (0.56) \\ 
1opk &   96.88 (0.05) & -224.05 (0.06) & -310.66 (0.48) &   98.21 (0.43) &  -85.29 (0.59) \\ 
1nav &   38.05 (0.02) &  -56.39 (0.05) & -124.70 (0.15) &   64.55 (0.68) &  -41.81 (0.64) \\ 
1xoz &   64.93 (0.07) &   63.37 (0.05) &  -12.67 (0.71) &   79.27 (0.36) &  -61.70 (0.67) \\ 
1n46 &   42.40 (0.04) &   -9.92 (0.06) &  -86.88 (0.23) &   39.61 (0.80) &  -79.74 (0.78) \\ 
1sg0 &   33.31 (0.01) &  -56.23 (0.04) & -108.03 (0.46) &   49.50 (0.70) &  -35.61 (0.82) \\ 
1z95 &  110.53 (0.05) &  -75.65 (0.07) & -133.16 (0.14) &  118.56 (0.93) &  -49.48 (0.86) \\ 
1m2z &   59.30 (0.07) &  104.73 (0.04) &   27.74 (0.07) &   64.66 (0.87) &  -71.63 (0.86) \\ 
1lpz &   45.72 (0.14) &  -23.41 (0.04) &  -93.41 (0.59) &   70.75 (0.85) &  -44.98 (0.97) \\ 
1t46 &  117.99 (0.06) & -172.73 (0.09) & -267.42 (0.53) &  109.84 (1.07) & -102.84 (1.09) \\ 
1gpk &   36.57 (0.05) &  -77.85 (0.05) & -155.19 (0.09) &   75.55 (1.20) &  -38.36 (1.16) \\ 
1ke5 &  127.49 (0.07) & -148.73 (0.04) & -197.35 (0.60) &  146.08 (1.05) &  -30.02 (1.20) \\ 
1r9o &   30.13 (0.03) &  -41.66 (0.05) &  -72.00 (0.33) &   23.62 (1.30) &  -36.85 (1.23) \\ 
2br1 &   90.52 (0.07) & -149.49 (0.07) & -202.12 (0.64) &  114.21 (1.07) &  -28.94 (1.23) \\ 
1w1p &   46.67 (0.15) &   30.21 (0.04) &    7.14 (0.11) &   57.48 (1.32) &  -12.26 (1.29) \\ 
1ywr &  102.56 (0.14) & -153.29 (0.06) & -214.19 (0.65) &   99.09 (1.01) &  -64.37 (1.47) \\ 
1q41 &   58.35 (0.07) &   16.62 (0.05) &  -30.16 (0.30) &   63.14 (1.33) &  -41.98 (1.47) \\ 
1ig3 &   70.96 (0.04) & -227.83 (0.06) & -285.08 (0.55) &  101.46 (1.38) &  -26.76 (1.51) \\ 
1owe &   34.40 (0.14) &  -47.49 (0.05) & -100.87 (0.23) &   46.79 (1.56) &  -40.99 (1.60) \\ 
1of1 &   70.34 (0.05) & -110.76 (0.07) & -160.84 (0.43) &   61.68 (1.59) &  -58.73 (1.62) \\ 
1kzk &   97.89 (0.17) &   14.29 (0.09) &  -74.63 (0.32) &  142.31 (1.60) &  -44.49 (1.67) \\ 
1v48 &   73.84 (0.05) & -192.27 (0.06) & -269.89 (1.30) &   76.26 (1.86) &  -75.21 (1.73) \\ 
2bm2 &   52.94 (0.06) &  -23.38 (0.12) & -104.05 (1.10) &   69.51 (0.95) &  -64.10 (1.73) \\ 
1n1m &   28.73 (0.09) &   27.22 (0.05) &   -6.80 (0.57) &    1.71 (1.84) &  -61.04 (1.76) \\ 
1q4g &   29.52 (0.02) &  -39.08 (0.04) &  -90.48 (0.11) &   30.71 (1.75) &  -50.20 (1.77) \\ 
1pmn &   80.90 (0.05) & -107.45 (0.08) & -173.96 (1.29) &   84.39 (0.69) &  -63.01 (1.82) \\
1w2g &   76.79 (0.24) & -132.10 (0.07) & -191.97 (0.18) &   78.18 (1.69) &  -58.49 (1.88) \\
1tow &   36.51 (0.03) &  -41.38 (0.04) &  -99.13 (0.30) &   76.82 (2.09) &  -17.44 (2.01) \\ 
1meh &   61.49 (0.05) &  -65.14 (0.05) & -120.95 (0.75) &   92.35 (1.52) &  -24.95 (2.07) \\ 
1hvy &   86.43 (0.10) & -214.15 (0.04) & -320.45 (0.82) &  134.06 (2.62) &  -58.67 (2.12) \\ 
1ygc &  150.51 (0.09) & -245.11 (0.07) & -330.67 (0.73) &  183.37 (2.16) &  -52.71 (2.17) \\ 
1tt1 &   82.99 (0.03) &  -57.77 (0.06) &  -94.79 (0.15) &   57.39 (2.24) &  -62.62 (2.18) \\ 
1s3v &  100.87 (0.08) & -255.96 (0.07) & -321.24 (1.23) &  104.11 (2.12) &  -62.04 (2.25) \\ 
1k3u &   50.03 (0.08) & -116.36 (0.03) & -169.77 (0.17) &   56.16 (2.34) &  -47.30 (2.34) \\ 
1oyt &   63.82 (0.17) &  -73.57 (0.12) & -142.23 (2.00) &   86.02 (2.40) &  -46.45 (2.41) \\ 
1n2v &   56.64 (0.05) &   42.14 (0.03) &    1.50 (0.39) &   69.01 (2.52) &  -28.28 (2.48) \\ 
1y6b &  118.26 (0.05) & -115.82 (0.08) & -180.78 (0.72) &  132.17 (2.38) &  -51.05 (2.51) \\ 
2bsm &   55.96 (0.05) &    3.86 (0.05) &  -45.97 (1.75) &   65.16 (1.97) &  -40.63 (2.53) \\ 
1vcj &  111.44 (0.16) &  -59.31 (0.08) & -128.13 (1.58) &  137.95 (2.47) &  -42.30 (2.57) \\ 
1unl &  133.75 (0.08) & -295.42 (0.08) & -347.74 (1.93) &  157.99 (1.54) &  -28.08 (2.58) \\ 
1ia1 &   83.97 (0.09) & -252.59 (0.04) & -288.51 (0.70) &  100.56 (2.37) &  -19.32 (2.58) \\ 
1gm8 &   78.20 (0.17) &  -30.20 (0.05) &  -84.37 (1.61) &  103.98 (2.10) &  -28.39 (2.62) \\ 
1hnn &   91.00 (0.01) & -139.60 (0.04) & -216.64 (0.21) &  143.38 (2.57) &  -24.66 (2.63) \\ 
1s19 &   39.46 (0.01) &   65.53 (0.12) &   -3.81 (0.23) &   25.23 (2.53) &  -83.57 (2.63) \\ 
1xoq &   54.13 (0.07) &   -1.10 (0.05) &  -60.98 (1.49) &   96.98 (1.96) &  -17.04 (2.63) \\ 
1l2s &  108.62 (0.03) & -159.94 (0.05) & -203.82 (0.28) &   93.68 (2.73) &  -58.82 (2.77) \\ 
1of6 &   72.62 (0.02) &  -37.28 (0.06) &  -68.64 (0.07) &   37.85 (2.79) &  -66.12 (2.82) \\ 
1g9v &   53.63 (0.05) &  -35.00 (0.04) &  -85.35 (0.14) &   89.92 (2.79) &  -14.06 (2.84) \\ 
1n2j &   44.11 (0.01) &  -12.59 (0.06) &  -51.33 (0.67) &   66.79 (2.70) &  -16.06 (2.92) \\ 
1hq2 &  103.12 (0.12) & -221.27 (0.03) & -267.22 (0.14) &  119.18 (2.96) &  -29.89 (2.94) \\ 
1j3j &   80.28 (0.10) & -246.45 (0.04) & -296.37 (0.32) &  100.14 (2.94) &  -30.06 (2.96) \\ 
1mzc &   59.99 (0.07) &  -28.41 (0.09) &  -94.13 (1.02) &   62.75 (2.76) &  -62.96 (2.97) \\ 
1t9b &  224.57 (0.03) & -701.13 (0.06) & -761.78 (2.33) &  245.95 (2.64) &  -39.27 (2.98) \\ 
1sj0 &   29.01 (0.04) &   39.51 (0.09) &  -57.97 (0.42) &   40.81 (3.17) &  -85.68 (3.02) \\ 
1v0p &  145.18 (0.09) & -331.87 (0.07) & -400.08 (1.56) &  169.28 (2.78) &  -44.11 (3.27) \\ 
1sq5 &   67.61 (0.07) &  -72.68 (0.08) & -109.64 (0.14) &  109.62 (3.60) &    5.05 (3.52) \\ 
1u4d &  104.66 (0.10) & -154.50 (0.05) & -193.28 (0.45) &  115.92 (3.60) &  -27.52 (3.59) \\ 
1gkc &   58.34 (0.14) &  -26.86 (0.08) &  -60.75 (1.20) &   69.69 (3.00) &  -22.53 (3.60) \\ 
1q1g &   64.44 (0.03) &  -66.36 (0.03) & -159.62 (0.31) &   96.52 (3.69) &  -61.18 (3.61) \\ 
1v4s &   86.02 (0.10) & -140.14 (0.03) & -200.32 (0.39) &  101.44 (3.99) &  -44.76 (3.96) \\ 
1jd0 &  117.54 (0.04) & -127.93 (0.04) & -175.59 (0.60) &  106.50 (3.73) &  -58.69 (4.14) \\ 
1lrh &   28.56 (0.02) &  -53.30 (0.05) & -147.43 (0.39) &   34.85 (4.26) &  -87.84 (4.22) \\ 
1u1c &   68.81 (0.06) & -123.44 (0.04) & -175.61 (0.90) &   64.80 (3.99) &  -56.18 (4.31) \\ 
1uml &   87.76 (0.08) &  -35.50 (0.08) & -100.68 (1.24) &  156.01 (4.73) &    3.07 (4.34) \\ 
1p62 &  101.68 (0.43) & -157.01 (0.06) & -190.81 (0.09) &  112.24 (4.28) &  -23.23 (4.40) \\ 
1t40 &   64.78 (0.06) &  -57.26 (0.05) & -148.56 (1.63) &   82.83 (4.11) &  -73.26 (5.27) \\ 
1mmv &   67.30 (0.05) &  -46.85 (0.06) & -131.27 (0.41) &  128.53 (5.49) &  -23.19 (5.31) \\ 
1hp0 &  106.74 (0.37) &  -89.17 (0.03) & -143.46 (0.47) &  148.46 (5.58) &  -12.57 (5.49) \\ 
1r58 &   73.47 (0.11) &   33.79 (0.05) &  -22.84 (1.69) &  132.55 (5.39) &    2.46 (5.51) \\ 
1jje &   26.48 (0.02) &  -96.77 (0.05) & -310.91 (1.17) &  160.62 (5.31) &  -80.00 (5.67) \\ 
1yqy &   91.67 (0.04) &  -80.06 (0.07) & -117.24 (0.77) &   60.96 (5.70) &  -67.89 (6.20) \\ 
1hwi &   61.72 (0.03) &  -75.12 (0.06) & -138.73 (1.22) &  111.60 (6.08) &  -13.74 (6.24) \\ 
1uou &   67.45 (0.08) & -159.93 (0.05) & -206.43 (0.12) &   79.48 (6.60) &  -34.47 (6.59) \\ 
1l7f &  106.44 (0.09) & -168.94 (0.10) & -231.83 (3.41) &   96.95 (6.12) &  -72.38 (7.10) \\ 
1tz8 &   23.80 (0.02) &   22.96 (0.04) &   19.54 (0.31) &   29.99 (7.43) &    2.76 (7.25) \\ 
1p2y &   16.48 (0.02) &   14.68 (0.03) &  -98.07 (0.07) &   93.47 (7.26) &  -35.76 (7.25) \\ 
1oq5 &  111.73 (0.03) &  -64.62 (0.05) & -125.02 (0.46) &   36.12 (7.43) & -136.00 (7.43) \\ 
1x8x &   72.62 (0.02) &  -37.29 (0.03) &  -70.80 (0.26) &   74.66 (8.10) &  -31.47 (8.11) \\ 
1r55 &   64.36 (0.11) &   -6.11 (0.15) &  -45.15 (2.03) &   83.98 (9.81) &  -19.43 (10.48) \\ 
1r1h &  132.41 (0.08) &  -71.65 (0.07) & -175.35 (0.49) &   79.09 (10.67) & -157.01 (10.61) \\ 
1xm6 &   39.40 (0.07) &    1.83 (0.05) &  -73.02 (2.08) &  263.42 (10.84) &  149.18 (11.10) \\ 
1hww &   26.16 (0.01) &   64.60 (0.04) &   24.37 (0.58) &   33.96 (11.12) &  -32.43 (11.19) \\ 
\hline
\caption{The mean (and, in parentheses, the standard deviation), of 11 independent free energy calculations for the complexes in the Astex Diverse set. Table entries are ordered by increasing standard deviation of the total BPMF estimate.}
\end{longtable}
\end{center}

\end{document}